\documentclass[twocolumn]{aastex3}

\usepackage{amsmath}
\usepackage{graphicx}
\usepackage{longtable}

\definecolor{myblue}{RGB}{0, 160, 240} 
\definecolor{mygreen}{RGB}{0, 180, 0}

\newcommand{\new}{\textcolor{black}}

\received{29 January 2021}
\revised{12 March 2021}
\accepted{18 March 2021}

\submitjournal{ApJ}

\shorttitle{The influence of metallicity on the Leavitt Law}
\shortauthors{Breuval et al.}

\begin{document}

\title{The influence of metallicity on the Leavitt Law \\ from geometrical distances of Milky Way and Magellanic Clouds Cepheids}

\author[0000-0003-3889-7709]{Louise Breuval}
\affiliation{LESIA, Observatoire de Paris, Universit\'e PSL, CNRS, Sorbonne Universit\'e, Universit\'e de Paris, 5 place Jules Janssen, 92195 Meudon, France}
\email{louise.breuval@obspm.fr}

\author[0000-0003-0626-1749]{Pierre Kervella}
\affiliation{LESIA, Observatoire de Paris, Universit\'e PSL, CNRS, Sorbonne Universit\'e, Universit\'e de Paris, 5 place Jules Janssen, 92195 Meudon, France}

\author[0000-0002-1662-5756]{Piotr Wielg\'orski}
\affiliation{Nicolaus Copernicus Astronomical Centre, Polish Academy of Sciences, Bartycka 18, 00-716 Warszawa, Poland}

\author[0000-0003-1405-9954]{Wolfgang Gieren} 
\affiliation{Universidad de Concepci\'on, Departamento de Astronom\'ia, Casilla 160-C, Concepci\'on, Chile }

\author[0000-0002-7355-9775]{Dariusz Graczyk}
\affiliation{Centrum Astronomiczne im. Miko\l aja Kopernika, Polish Academy of Sciences, Rabia\'nska 8, 87-100, Toru\'n, Poland}

\author[0000-0001-5875-5340]{Boris Trahin}
\affiliation{LESIA, Observatoire de Paris, Universit\'e PSL, CNRS, Sorbonne Universit\'e, Universit\'e de Paris, 5 place Jules Janssen, 92195 Meudon, France}

\author[0000-0002-9443-4138]{Grzegorz Pietrzy\'nski}
\affiliation{Nicolaus Copernicus Astronomical Centre, Polish Academy of Sciences, Bartycka 18, 00-716 Warszawa, Poland}

\author[0000-0003-2837-3899]{Fr\'ed\'eric Arenou}
\affiliation{GEPI, Observatoire de Paris, Universit\'e PSL, CNRS, 5 place Jules Janssen, 92190 Meudon, France}

\author[0000-0002-9317-6114]{Behnam Javanmardi}
\affiliation{LESIA, Observatoire de Paris, Universit\'e PSL, CNRS, Sorbonne Universit\'e, Universit\'e de Paris, 5 place Jules Janssen, 92195 Meudon, France}

\author[0000-0003-1515-6107]{Bart\l omiej Zgirski}
\affiliation{Nicolaus Copernicus Astronomical Centre, Polish Academy of Sciences, Bartycka 18, 00-716 Warszawa, Poland}


\begin{abstract}

The Cepheid Period-Luminosity (PL) relation is the key tool for measuring astronomical distances and for establishing the extragalactic distance scale. In particular, the local value of the Hubble constant ($H_0$) strongly depends on Cepheid distance measurements. The recent \textit{Gaia} Data Releases and other parallax measurements from the \textit{Hubble} Space Telescope (HST) already enabled to improve the accuracy of the slope ($\alpha$) and intercept ($\beta$) of the PL relation. However, the dependence of this law on metallicity is still largely debated. In this paper, we combine three samples of Cepheids in the Milky Way (MW), the Large Magellanic Cloud (LMC) and the Small Magellanic Cloud (SMC) in order to derive the metallicity term (hereafter $\gamma$) of the PL relation. 
The recent publication of extremely precise LMC and SMC distances based on late-type detached eclipsing binary systems (DEBs) provides a solid anchor for the Magellanic Clouds. In the MW, we adopt Cepheid parallaxes from the early third \textit{Gaia} Data Release. We derive the metallicity effect in $V$, $I$, $J$, $H$, $K_S$, $W_{VI}$ and $W_{JK}$. In the $K_S$ band we report a metallicity effect of $-0.221 \pm 0.051$ mag/dex, the negative sign meaning that more metal-rich Cepheids are intrinsically brighter than their more metal-poor counterparts of the same pulsation period.
\end{abstract}

\keywords{parallaxes -- stars: distances -- stars: variables: Cepheids -- distance scale -- metallicity}

\section{Introduction} 
\label{sec:intro}

The Cepheid Period-Luminosity (PL) relation, discovered by Henrietta Leavitt \citep{Leavitt1912} about a century ago, is an essential tool for measuring astronomical distances since it represents the first rung of the extragalactic distance ladder. This law is used to measure distances to type Ia supernovae \new{(SNe Ia)} host galaxies, and thus plays a key role in the determination of the Hubble constant ($H_0$). This parameter currently exhibits a tension of at least $\sim$ 4$\sigma$ between its measurement in the early Universe by \citet{Planck2018} assuming a $\Lambda$-CDM cosmology and the local estimate based on Cepheid distances \citep{Riess2021}. The precise calibration of the PL relation is therefore of paramount importance to reach a 1\% determination of the Hubble constant.

While the slope ($\alpha$) and intercept ($\beta$) of the Leavitt law are generally consistent between various studies, the value and even the sign of the metallicity term ($\gamma$, defined as $M=\alpha \log P + \beta + \gamma \rm [Fe/H]$) are still debated and constitute $0.5 \%$ of the error budget of $H_0$ \citep{Riess2016}. Some empirical studies report a metallicity dependence consistent with $\gamma \sim 0 ~ \rm mag/dex$: \citet{Udalski2001} concludes with a null-effect from the study of a metal-poor galaxy in optical bands, \citet{Storm2011b} finds a null effect in all bands except in $W_{VI}$ and \citet{Wielgorski2017} derive a gamma value consistent with zero in optical and NIR bands. Still, a large majority of the analysis investigating the metallicity effect derived a negative sign, with values ranging between $-0.2$ and $-0.5\, \rm mag/dex$ \citep{Freedman1990, Macri2006, Saha2006, Gieren2018, Groenewegen2018}. This trend would indicate that metal-rich Cepheids are brighter than metal-poor ones. However, the study by \citet{Romaniello2008} yielded a metallicity effect of the opposite sign, confirming the theoretical predictions \citep{Caputo2000, Bono2008, Fiorentino2013}. 

In this paper, we aim at determining the effect of metallicity on the PL relation by combining samples of Cepheids in the Milky Way (MW) and in the Magellanic Clouds (MCs), taking advantage of the large range of metallicity covered by the Cepheids in these 3 galaxies (from $+0.08$ dex to $-0.75$ dex). Most of the Cepheids located in distant galaxies hosting SNIa have metallicities within this range, therefore our results are directly applicable to extragalactic studies of the distance scale \citep[e.g.][]{Javanmardi2021}.

Recently, \citet{Pietrzynski2019} and \citet{Graczyk2020} measured the most precise distances to date for the Large Magellanic Cloud (LMC) and Small Magellanic Cloud (SMC) respectively, based on enhanced samples of late-type detached eclipsing binaries (DEBs). These distances allow us to obtain a precise calibration of the PL relation in the LMC and SMC. For Milky Way (MW) Cepheids, we use the early third \textit{Gaia} Data Release (EDR3) which recently provided parallaxes of unprecedented precision for hundreds of galactic Cepheids. 

In Sect. \ref{sect:sample}, we present our samples of Cepheids in the three galaxies and in Sect. \ref{sect:distances} we provide the distances we adopted for each sample. Then in Sect. \ref{sect:gamma_3D} we estimate the metallicity effect by fitting the Period-Luminosity-Metallicity (PLZ) relation in the three galaxies. We discuss the results in Sect. \ref{sect:discussion}.  \\

\begin{figure*}[]
\centering
\includegraphics[height=5.65cm]{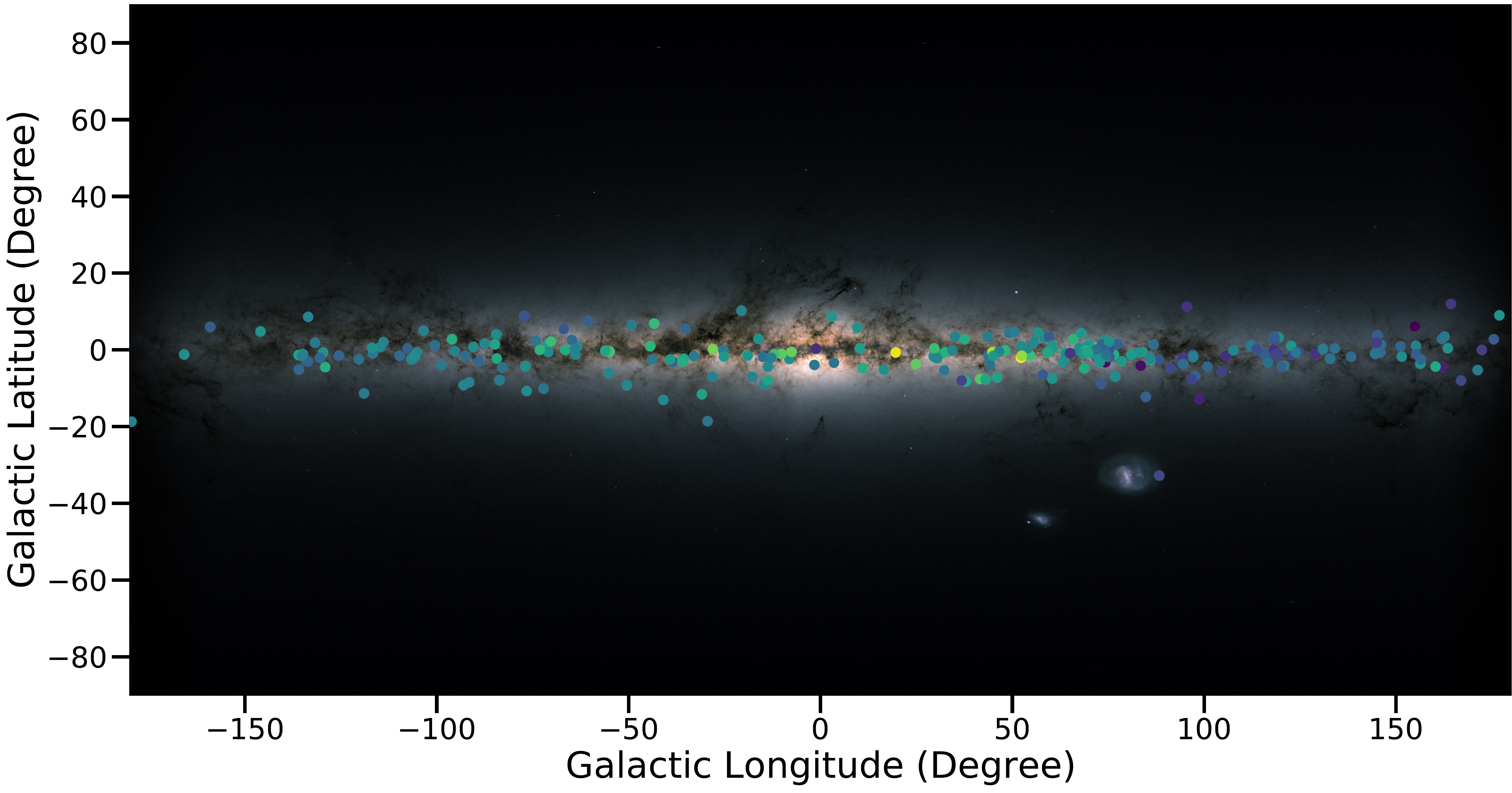}
\includegraphics[height=5.65cm]{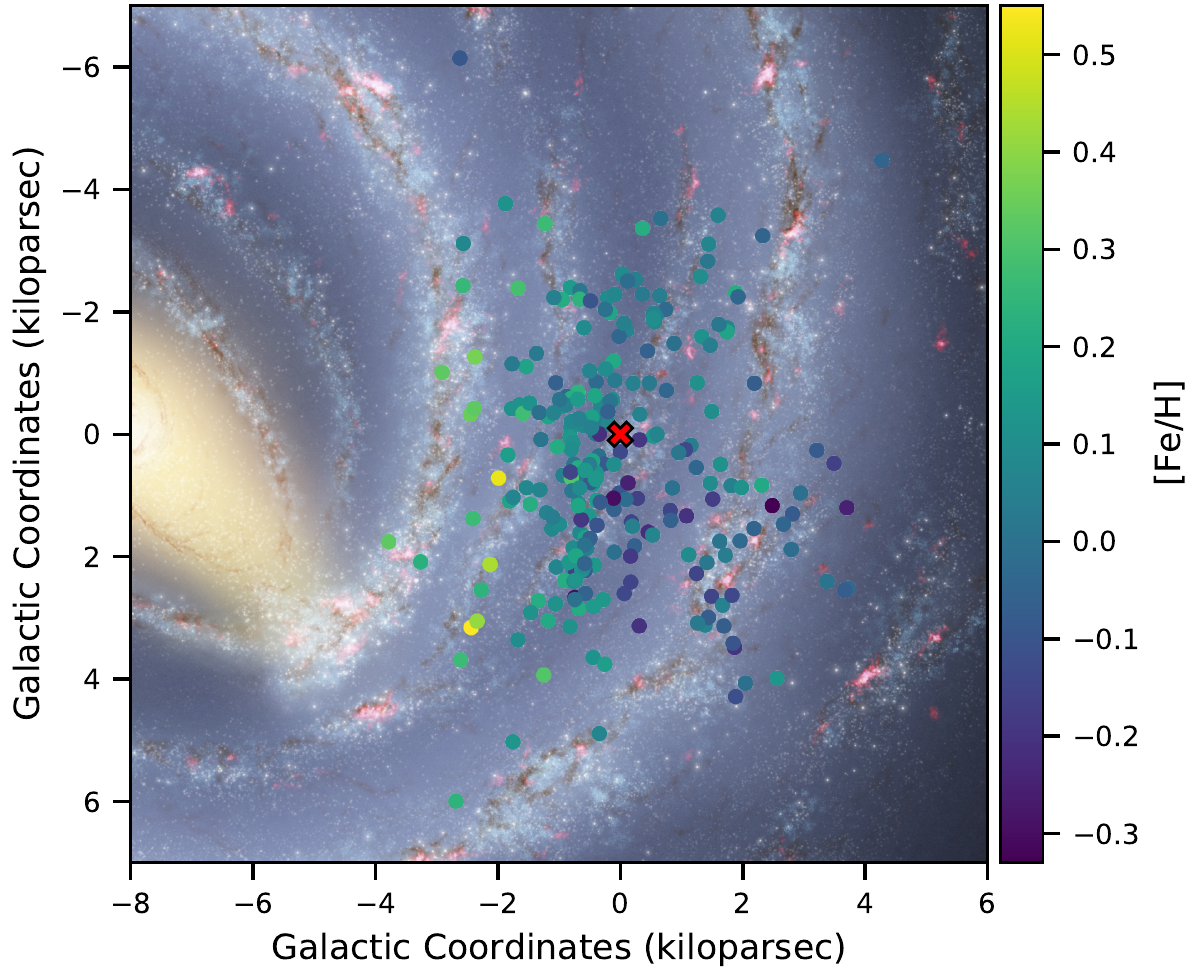}
\caption{Galactic maps projected on the sky (left) and on the galactic plane (right) showing the distribution of the MW Cepheid sample. The color scale represents the metallicity [Fe/H] and the red cross is the position of the Solar System.}
\label{fig:map_MW}
\end{figure*}

\section{Samples of Cepheids} 
\label{sect:sample}

        \subsection{MW Cepheids}
        \label{subsec:MW}
        
We gather a sample of Milky Way Cepheids for which well-covered light curves are available. In the NIR $J$, $H$ and $K$ bands, we combine the catalogs by \citet{Welch1984}, \citet{Laney1992}, \citet{Barnes1997} and \citet{Monson2011}. The data from these four studies are found to be in close agreement, with residuals of 0.013, 0.010 and 0.002 mag in $J$, $H$ and $K$ respectively \citep{Monson2011}. We adopt these values as photometric zero-point uncertainties for the NIR photometry. Additional NIR data were also found in \citet{Feast2008}, we consider that including this source of data does not impact the homogeneity and the dispersion of the data since it only affects four stars of the sample. In the optical $V$ and $I$ bands, we use the catalog from \citet{Berdnikov2008} that provides photometry in the Johnson-Cousins system for a large number of Cepheids. Since it is a compilation of data from various catalogs by the same author, we adopt a photometric zero-point uncertainty of 0.010 mag.

For each star and in each filter, we phase the data at the date of maximum luminosity and we obtain intensity-averaged mean apparent magnitudes by performing light curve fitting using Fourier series. Depending on the properties of the different light curves (such as the presence of bumps, steep variations, or to prevent the introduction of unphysical oscillations when the data are too dispersed or not dense enough), we adapt the number of Fourier modes, and thus of free parameters, in order to obtain a satisfactory representation of the light curve. A Fourier decomposition of order three is generally sufficient for an usual Cepheid light curve such as $\delta$ Cep, and is up to order six for a more complex star such as RS Pup. We derived the statistical uncertainties on the mean magnitudes from the scatter of each light curve. In some few cases, a very large number of data points are available ($>$ 300) and result in unrealistic small errors: in these cases we adopt a minimum error of 0.006 mag.

For long period Cepheids, large phase shifts may degrade the quality of the fit, the photometry being spread over four decades. Therefore, period changes were taken into account for the phasing of long period stars such as SV Vul, GY Sge or RS Pup \citep{Kervella2017}. We adopted a polynomial model of up to degree five for the pulsation period.

We carefully analyse the light curves: we exclude Cepheids for which less than 8 data-points are available (MW Cepheids have on average 35 data points in NIR and 160 in optical) and Cepheids that have poor quality photometry or insufficient phase coverage. Finally, we convert all the NIR data in the 2MASS system using the transformations from \citet{Monson2011}.  The systematics related to these transformations are negligible. Examples of a well covered light curve and of a poor-quality light curve are provided in Fig. \ref{fig:good_LC} and \ref{fig:bad_LC} in Appendix. 

We select Cepheids pulsating in the fundamental mode according to the reclassification by \citet{Ripepi2019}. For stars that were not available in this catalog, we adopted by order of priority the pulsation modes from \citet{Groenewegen2018}, from the Variable Star indeX \citep[VSX, ][]{VSX2006} and from \citet{Luck2018}. 

We adopt reddening values from \citet{Fernie1995} with a 0.94 scaling factor as suggested by \citet{Groenewegen2018}, and from \citet{Acharova2012} if not available in the latter. We adopt an uncertainty of 0.05 if it is not provided.

For MW Cepheids, we \new{search} for individual metallicities in \citet{Genovali2015}. This catalog provides mean abundances based on high resolution spectra for 75 Cepheids. For stars that are not available in this catalog, we adopt the values from \citet{Genovali2014}: they provide homogeneous Cepheid metallicities from their group and compiled from the literature, rescaled to their solar abundance. The individual metallicities are represented in Fig. \ref{fig:map_MW} by colored points, they range from $-0.33 \, \rm dex$ to $+0.55  \, \rm dex$. The gradient of metallicity in the MW is particularly visible, with metal-rich Cepheids located closer to the galactic center than metal-poor ones. These individual metallicities have a weighted mean value of $0.083 \pm 0.019\, \rm dex$ with a scatter of 0.14 dex. In the following, we adopt this weighted mean value for all MW Cepheids for consistency and homogenity with the LMC and SMC samples that only have a mean metallicity, but also because the current precision of the individual metallicities is not sufficient for a thorough calibration of the metallicity effect. 

The Cepheids of our MW sample are represented in Fig. \ref{fig:map_MW} and their main parameters are listed in Table \ref{table:data_MW_parameters} and Table \ref{table:data_MW_photometry} in \new{Appendix}.   \\

        \subsection{LMC Cepheids}
        \label{subsec:LMC}

We build a sample of LMC Cepheids by combining the OGLE-IV photometry in $V$ and $I$ bands \citep{Soszynski2015} with the multi-epoch observations from the LMC Near-Infrared Synoptic Survey by \citet{Macri2015} taken with the CPAPIR camera on the 1.5m CTIO telescope. We update their NIR mean magnitudes to bring them into better agreement with the 2MASS system using the following relations (L. Macri, priv. comm.). These were derived by matching $\sim$ 34 000 stars in common between their Table A1 and the 2MASS Point Source Catalog \citep{Cutri2003}, with $12<H<13.5$, $K>11.5$ and $-0.5<J-K<1.4$ mag:
\[
\begin{array}{l l l l l l l l}
J_{\rm 2MASS} = J_{\rm M15} - 0.0167 + 0.0205 \, (J_{\rm M15}-K_{\rm M15}-0.4) \\ 
\hspace{1.6cm} + 0.0101\, (J_{\rm M15}-K_{\rm M15}-0.4)^2 \\
H_{\rm 2MASS} = H_{\rm M15} + 0.0116 - 0.0054 \, (J_{\rm M15}-K_{\rm M15}-0.4) \\
\hspace{1.6cm} - 0.0189 \, (J_{\rm M15}-K_{\rm M15}-0.4)^2 \\
K_{\rm 2MASS} = K_{\rm M15} + 0.0162 + 0.0227 \, (J_{\rm M15}-K_{\rm M15}-0.4) \\
\hspace{1.6cm}- 0.0595 \, (J_{\rm M15}-K_{\rm M15}-0.4)^2
\end{array}
\]
We adopt a photometric zero-point uncertainty of 0.02 mag in all bands.
Since some Cepheids exhibit large brightness variations during a pulsation cycle, we consider that single-epoch photometry is not precise enough to derive reliable mean magnitudes, therefore we discarded the mean magnitudes derived by \citet{Inno2016} from template fitting on 2MASS single-point data and IRSF measurements. 

\begin{figure}[]
\centering
\includegraphics[height=7.2cm]{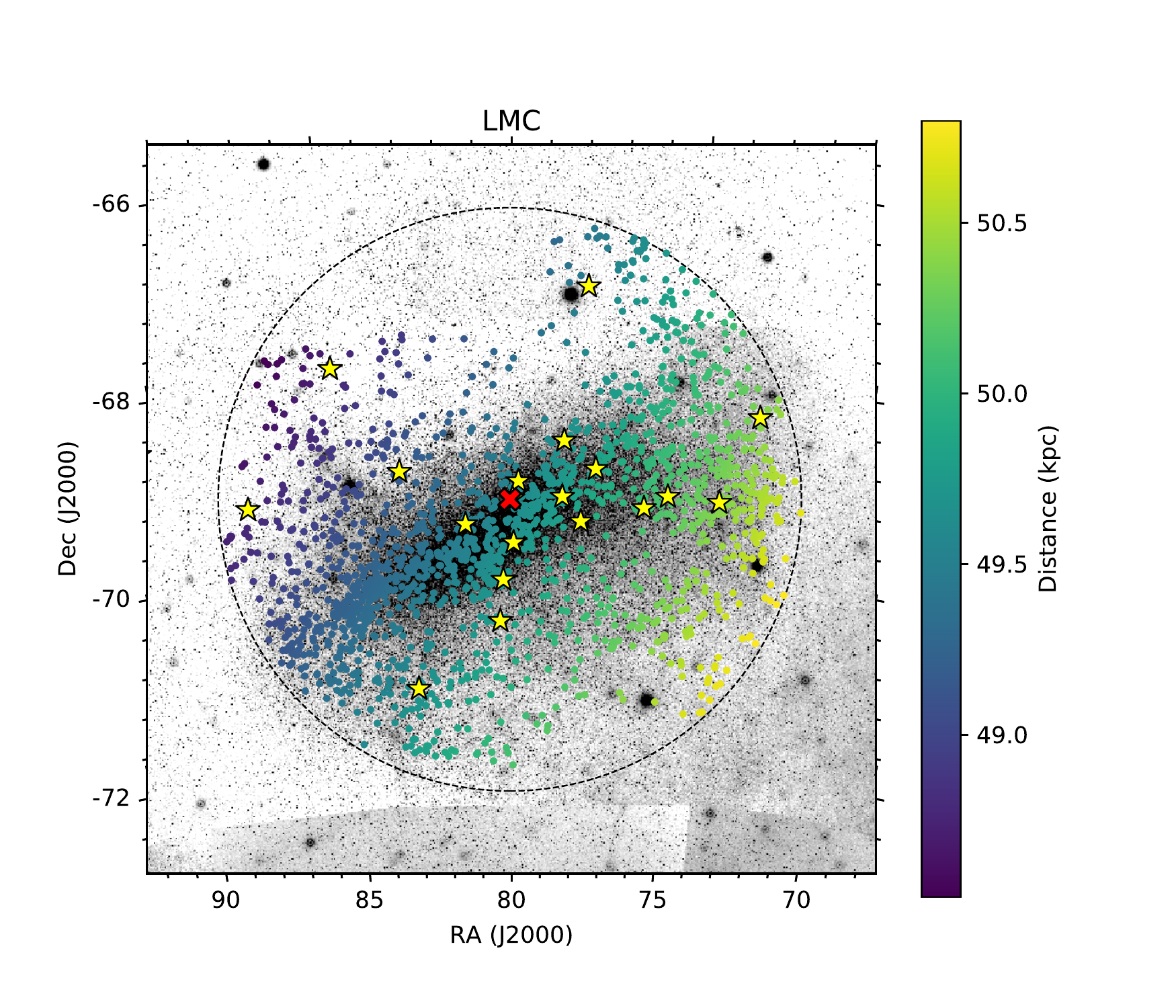} 
\caption{Map of the LMC Cepheids considered in our study. Yellow stars are the eclipsing binaries from \citet{Pietrzynski2019} and the red cross is the center of the LMC. The dashed circle represents a radius of 3 degrees around the LMC center.}
\label{fig:map_LMC}
\end{figure}

We perform a quality check on this initial sample: we reject stars with magnitude uncertainties larger than 1\% and with less than 5 data points (LMC Cepheids have on average 43 data points in NIR and 147 in optical), and we only consider fundamental mode Cepheids. We reject Cepheids located outside a radius of 3 degrees around the LMC center in order to avoid outliers such as stars that do not belong to the LMC or that are strongly affected by its geometrical effects (see Sect. \ref{sect:dist_LMC}). We adopt reddening values from the \citet{Gorski2020} reddening map. The final sample of LMC Cepheids contains 1446 stars in the $V$ band and 807 stars in $K_S$, \new{it is listed in Table \ref{table:data_LMC} in Appendix and provided as supplementary material online}. A map of the final sample of LMC Cepheids is represented in Fig. \ref{fig:map_LMC}. For LMC Cepheids, we adopt the mean metallicity used by \citet{Gieren2018}, that compiles several estimates from various studies: $[\rm Fe/H]_{\rm LMC} = -0.34 \pm 0.06 \, \rm dex$. The uncertainties take into account the homogenization of the different measurements.   \\

        \subsection{SMC Cepheids}
        \label{subsec:SMC}

We assemble a sample of SMC Cepheids by taking the mean magnitudes from the VISTA survey of the Magellanic Clouds (VMC) \citep{Ripepi2016} cross-matched with OGLE IV photometry by \citet{Soszynski2015}. Unfortunately, we do not have $H$-band photometry for SMC Cepheids because we rejected data from single epoch photometry and template fitting. Results in the $H$ band are therefore derived from the combination of MW and LMC Cepheids only. Magnitudes in the VISTA system were converted into the 2MASS system using the equations from \citet{Ripepi2016}:
\[
\begin{array}{l l l l l l l l}
J' &= &J_{\rm VMC} &+ &0.070 \, &(J_{\rm VMC}-K_{\rm VMC}) \\ 
K' &= &K_{\rm VMC} &- &0.011 \, &(J_{\rm VMC}-K_{\rm VMC}) \\ 
\end{array}
\]
We perform an additional correction (L. Macri, priv. comm.) derived by matching $\sim$ 7000 stars in common between the VMC DR4 and the 2MASS Point Source Catalog, with $J>12.25$, $K>11.5$ and $-0.5<J-K<1.4$ mag:
\[
\begin{array}{l l l l l l l l}
J_{\rm 2MASS} &= &J'  &- &0.0087 &-& 0.0010 &(J'-K'-0.4) \\ 
K_{\rm 2MASS} &= &K' &+ &0.0011 &-& 0.0087&(J'-K'-0.4) \\
\end{array}
\]
We adopt a photometric zero-point uncertainty of 0.02 mag for all bands. As we did for the LMC sample, we also reject SMC Cepheids with magnitude uncertainties larger than 1\%, with less than 5 data points (SMC Cepheids have on average 17 data points in NIR and 46 in optical) and we only keep Cepheids pulsating in the fundamental mode. As for the LMC sample, we adopt reddening values from the \citet{Gorski2020} reddening map.

\begin{figure}[]
\centering
\includegraphics[height=7.1cm]{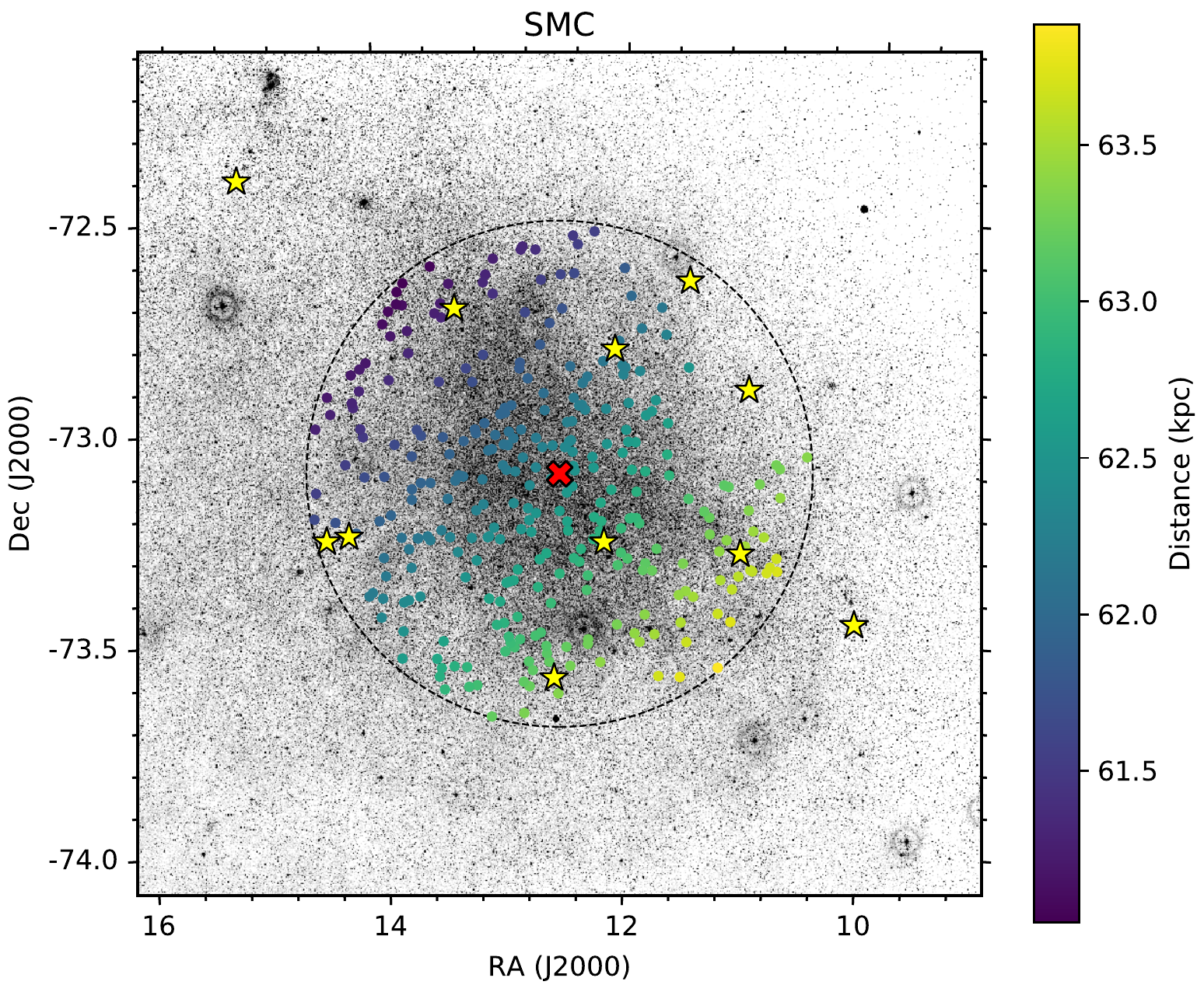}
\caption{Map of the SMC Cepheids considered in our study. Yellow stars are the eclipsing binaries from \citet{Graczyk2020} and the red cross is the center of the SMC. The dashed circle represents a radius of 0.6 degree around the SMC center.}
\label{fig:map_SMC}
\end{figure}

While the LMC has a rather simple geometry, the SMC is very elongated along the line of sight: we select Cepheids located in a region of 0.6 deg around the SMC center, which covers an area of 1.3 kpc width. Since the SMC distance is derived from detached eclipsing binaries (DEBs), this selection ensures that the Cepheids are located in the same region as these DEBs. The final SMC sample has 284 stars in the $V$ band and 295 stars in $K_S$, \new{it is listed in Table \ref{table:data_SMC} in Appendix and provided as supplementary material}. A map of our final sample of SMC Cepheids is represented in Fig. \ref{fig:map_SMC}. 

For SMC Cepheids, we adopt the mean metallicity used by \citet{Gieren2018}, that compiles several estimates from various studies: $[\rm Fe/H]_{\rm SMC} = -0.75 \pm 0.05 \, \rm dex$. Similar to the LMC value, the uncertainty takes into account the homogenization of the different measurements.   \\

\section{Distances} 
\label{sect:distances}

In order to calibrate the Leavitt law, one needs to derive the absolute magnitude of each Cepheid from its apparent luminosity and from its distance. \\

        \subsection{Distances to Milky Way Cepheids}
        \label{sect:EDR3}
        
Recently, the early third \textit{Gaia} Data Release provided new parallaxes for Milky Way Cepheids \citep{GaiaEDR3}. We perform a first quality check of \textit{Gaia} EDR3 parallaxes based on the Renormalised Unit Weight Error (RUWE) provided in the catalog. This parameter reflects the quality of the parallax of a star compared to other stars of the same color and brightness. Its value is expected to be close to 1 for well-behaved sources \citep{Lindegren2020a}. In particular, the RUWE is sensitive to the photocentric motion of unresolved objects, therefore it can be used to detect possible astrometric binaries. We discard the Cepheids of our sample that have a $\rm RUWE > 1.4$: this selection corresponds to approximately 13 \% of our MW sample and removes the stars that are possibly affected by saturation or contamination by a bright neighbour companion. In particular, all the outliers noticed by eye on the PL relation are affected by a large RUWE, therefore the quality check based on this parameter appears to be relevant for our purpose.

\citet{Riess2021} use a different indicator: they identify stars with a goodness of fit (GOF) larger than 12.5 as having a compromised parallax. We find the GOF and the RUWE selections to have a very similar effect on our sample: adopting this GOF criteria for the quality check instead of the RUWE leads to rejecting exactly the same stars, except T Mon, V0496 Aql and VW Pup, that have a RUWE of 1.72, 1.56 and 1.41, and a GOF of 12.11, 10.90 and 11.36 respectively. The RUWE criterion seems slightly more selective than the GOF limit adopted by \citet{Riess2021}. Adopting a threshold of RUWE $<$ 1.4 corresponds to a limit GOF of 10.

One method to check if a Cepheid is an astrometric binary is to look for proper motion anomalies between the observations by \textit{Hipparcos} and \textit{Gaia}. Using the approach described in \citet{Kervella2019a}, we find 20 Cepheids with a high proper motion anomaly signal. However, none of them were identified by their high RUWE or GOF and they do not appear as outliers, therefore we do not exclude them.

Cepheids are variable stars and therefore their brightness and colour can change significantly during a pulsation cycle. This effect was not taken into account in the processing of \textit{Gaia} DR2 astrometry and resulted in additional systematics, noise and dispersion for variable stars parallaxes \citep{Breuval2020}. The correction for this chromatic effect on Cepheid parallaxes is still absent from \textit{Gaia} EDR3 \citep{Lindegren2020a}. However, the number of observations obtained for each star increased consequently between \textit{Gaia} DR2 ($\sim$22 months) and \textit{Gaia} EDR3 ($\sim$34 months). We assume in this paper that the noise induced by this effect is negligible for \textit{Gaia} EDR3 parallaxes.

For each Cepheid we correct for the parallax zero-point (ZP) by using the Python tool\footnote{\url{https://www.cosmos.esa.int/web/gaia/edr3-code}} described in \citet{Lindegren2020a}. This ZP correction takes into account the ecliptic latitude, magnitude and colour of each star. Our MW Cepheids cover a range of magnitudes from $G= 3$ to $G= 12$ mag. For our sample of MW Cepheids, we find the ZP to vary between $-4$ and $-54 \, \rm \mu as$ with a median value of $-27 \, \rm \mu as$ ($\sigma = 10 \, \rm \mu as$), which is very similar to the median parallax offset derived by \citet{Riess2021}. Following \citet{Lindegren2020a} who recommend to include an uncertainty of a few micro arcseconds in the ZP, we adopt a systematic error of \new{5~$\mu$as} on this quantity. Considering our sample of Cepheids, this error is equivalent to an average systematic uncertainty of 0.020 mag in distance modulus. In Sect. \ref{sect:ZP}, we discuss the influence of adopting this individual ZP correction compared with the uniform ZP of $-17 \, \rm \mu as$ derived from quasars. 

We find 13 Cepheids to fall in the range between $G= 10.8$ and $G= 11.2$ mag, where a transition of window classes occurs \citep[Fig. 1 in][]{Lindegren2020b}. In this particular range, the value of the parallax zero-point can possibly be affected so we quadratically add $10 \, \rm \mu as$ to the parallax uncertainty.

Finally, we increase all \textit{Gaia} EDR3 parallax uncertainties by 10\%, following \citet{Riess2021} to account for potential excess uncertainty. This correction has significantly reduced since \textit{Gaia} DR2, where it was recommended to increase parallax uncertainties by 30\%.\\

        \subsection{Distances to LMC Cepheids}
        \label{sect:dist_LMC}
        
Recently, \citet{Pietrzynski2019} estimated the distance to the LMC with a 1\% precision based on detached-eclipsing binaries (DEBs): $d_{\rm LMC} = 49.59 \pm 0.09 \rm \, (stat.) \pm 0.54 \rm \, (syst.)$ kpc. This method for measuring distances is independent from Cepheids and relies on surface-brightness relations, established by precise interferometric measurements. We use this value as initial distance to our Cepheids and we add a corrective term depending on the position of each Cepheid in the LMC, assuming the disc geometry derived by OGLE from Cepheids by \citet{Jacyszyn2016}. First we compute the cartesian coordinates $(x_i, ~ y_i, ~ z_i)$ of each Cepheid from their equatorial coordinates $(\rm \alpha_i, \, \rm \delta_i)$ using the transformations:
\[
\left\{
\begin{array}{l l l l l l l l}
x_i = -d_{\rm LMC}  \cos \delta_i  \sin(\rm \alpha_i - \rm \alpha_{\rm LMC}) \\
y_i = d_{\rm LMC} ~ \big[ \sin \delta_i  \cos \delta_{\rm LMC} \\
 \hspace{1.2cm} - \cos \delta_i  \sin \delta_{\rm LMC}  \cos(\rm \alpha_i - \rm \alpha_{\rm LMC}) \big] \\
z_i = c_1 x_i + c_2 y_i + d_{\rm LMC}
\end{array}
\right.
\]
where ($\alpha_{\rm LMC}$, $\delta_{\rm LMC}$) = (80.05, -69.30) deg are the coordinates of the LMC center and the coefficients $(c_1, c_2)$ = $(0.395 \pm 0.014,  -0.215 \pm 0.013)$ are from \citet{Jacyszyn2016}. The corrected distance of each LMC Cepheid is:
\[
\begin{array}{l l l l l l l l}
d_i = \sqrt {x_i^2 + y_i^2 + z_i^2}
\end{array}
\]
The distances of each LMC Cepheid derived with this correction are located in a range of $\pm$ 1.5 kpc around the mean LMC distance from \citet{Pietrzynski2019}. They are represented by the colors in Fig. \ref{fig:map_LMC}. \\

\begin{table}[]
\caption{Results of the PL fit of the form $M = \alpha (\log P - 0.7) + \beta$ in the Milky Way, the Large Magellanic Cloud and the Small Magellanic Cloud.}
\centering
\begin{tabular}{l c c c c c c l}
\hline
\hline
Band 	& $\alpha$ 			& $\beta$ 				&  $\sigma$ & $N^{(*)}$	 \\
\hline
\multicolumn{5}{c}{MW $^{(a)}$}  \\
\hline
 $V$  		& $-2.443 \pm 0.031$ & $-3.296 \pm 0.024$ & 0.25 & 178  \\ 
 $I$  			& $-2.780 \pm 0.028$ & $-3.981 \pm 0.024$ & 0.23 & 150  \\ 
 $W_{VI}$		& $-3.289 \pm 0.026$ & $-5.030 \pm 0.025$ & 0.21 & 149  \\ 
 $J$  		& $-3.050 \pm 0.029$ & $-4.498 \pm 0.026$ & 0.18 & 97   \\ 
 $H$			& $-3.160 \pm 0.028$ & $-4.762 \pm 0.024$ & 0.17 & 97   \\
 $K_S$ 		& $-3.207 \pm 0.028$ & $-4.848 \pm 0.022$ & 0.17 & 97  \\ 
 $W_{JK}$  	& $-3.317 \pm 0.028$ & $-5.086 \pm 0.026$ & 0.17 & 97  \\
\hline
\multicolumn{5}{c}{LMC  $^{(b)}$}  \\
\hline
 $V$  		& $-2.704 \pm 0.007$ & $-3.284 \pm 0.033$ & 0.23 & 1446  \\ 
 $I$  			& $-2.916 \pm 0.005$ & $-3.910 \pm 0.033$ & 0.15 & 1460  \\ 
 $W_{VI}$	  	& $-3.281 \pm 0.008$ & $-4.877 \pm 0.038$ & 0.08 & 1432  \\ 
 $J$  		& $-3.127 \pm 0.005$ & $-4.385 \pm 0.033$ & 0.12 & 805  \\ 
 $H$			& $-3.160 \pm 0.005$ & $-4.696 \pm 0.033$ & 0.11 & 808  \\
 $K_S$  		& $-3.217 \pm 0.005$ & $-4.737 \pm 0.033$ & 0.10 & 807  \\ 
 $W_{JK}$  	& $-3.272 \pm 0.008$ & $-4.974 \pm 0.039$ & 0.10 & 806  \\ 
\hline
\multicolumn{5}{c}{SMC  $^{(c)}$}  \\
\hline
 $V$  		& $-2.594 \pm 0.012$ & $-3.196 \pm 0.038$ & 0.28 & 284  \\ 
 $I$  			& $-2.871 \pm 0.008$ & $-3.841 \pm 0.038$ & 0.22 & 297  \\ 
 $W_{VI}$	  	& $-3.334 \pm 0.014$ & $-4.834 \pm 0.043$ & 0.12 & 283  \\ 
 $J$  		& $-2.956 \pm 0.004$ & $-4.317 \pm 0.038$ & 0.17 & 294   \\ 
 $H$  		&  ---				  &  --- 			    &  ---    & ---    \\ 
 $K_S$  		& $-3.163 \pm 0.002$ & $-4.670 \pm 0.038$ & 0.15 & 295  \\ 
 $W_{JK}$  	& $-3.326 \pm 0.002$ & $-4.916 \pm 0.043$ & 0.14 & 295  \\ 
\hline
\end{tabular}
\flushleft
\tablecomments{(*) The number of stars is given after the sigma clipping procedure and the period cuts. \\
(a) Mean [Fe/H] $= +0.083 \pm 0.019$ dex \\
(b) Mean [Fe/H] $= -0.34 \pm 0.06$ dex \\
(c) Mean [Fe/H] $= -0.75 \pm 0.05$ dex \\
}
\label{table:PL}
\end{table}

        \subsection{Distances to SMC Cepheids}
        \label{sect:dist_SMC}
        
The distance to the SMC was recently measured by \citet{Graczyk2020} with a precision of 1.5 \% using the same method as used in  \citet{Pietrzynski2019} for the LMC: from a sample of 15 DEBs, a distance of $d_{\rm SMC} = 62.44 \pm 0.47 \rm \, (stat.) \pm 0.81 \rm \, (syst.) $ kpc is derived. However, the SMC has a large extension along the line of sight \citep{Subramanian2012, Jacyszyn2016, Ripepi2017}, which makes the distance to its core region particularly difficult to measure, contrary to the LMC that has a rather simple geometry. In this section, we take into account the SMC elongated shape in order to derive corrected distances to each of its Cepheids. For each SMC Cepheid of coordinates $(\rm \alpha_i, \, \rm \delta_i)$, we compute the cartesian coordinates $(x_i, \, y_i)$ such that:
\[
\left\{
\begin{array}{l l l l l l l l}
x_i = -d_{\rm SMC}  \cos \delta_i  \sin(\rm \alpha_i - \rm \alpha_{\rm SMC}) \\
y_i = d_{\rm SMC} ~ \big[ \sin \delta_i  \cos \delta_{\rm SMC} \\
 \hspace{1.2cm} - \cos \delta_i  \sin \delta_{\rm SMC}  \cos(\rm \alpha_i - \rm \alpha_{\rm SMC}) \big] \\
\end{array}
\right.
\]
where ($\alpha_{\rm SMC}$, $\delta_{\rm SMC}$) = (12.54, -73.11) deg \citep{Ripepi2017}. Then we used the equations corresponding to the blue lines in Fig. 4 of \citet{Graczyk2020}:
\[
\left\{
\begin{array}{l l l l l l l l}
d_i(x) = (3.086 \pm 0.066) ~ x_i + d_{\rm SMC} \\
d_i(y) = (-3.248 \pm 0.118) ~ y_i + d_{\rm SMC} \\
\end{array}
\right.
\]
We adopt the mean value of $d_i(x)$ and $d_i(y)$ as the final distance of each Cepheid. The elongated shape of the SMC is highlighted by the dispersion of the derived distances between +5 kpc and -6 kpc around the mean value $d_{\rm SMC}$, which represents almost 10\% of the mean value. The distances of our sample of SMC Cepheids are represented on the map in Fig. \ref{fig:map_SMC}. A discussion about the elongated shape of the SMC and its impact on our results is provided in Sect. \ref{sect:influence_SMC}. 

\begin{figure*}[]
\centering
\includegraphics[height=10.5cm]{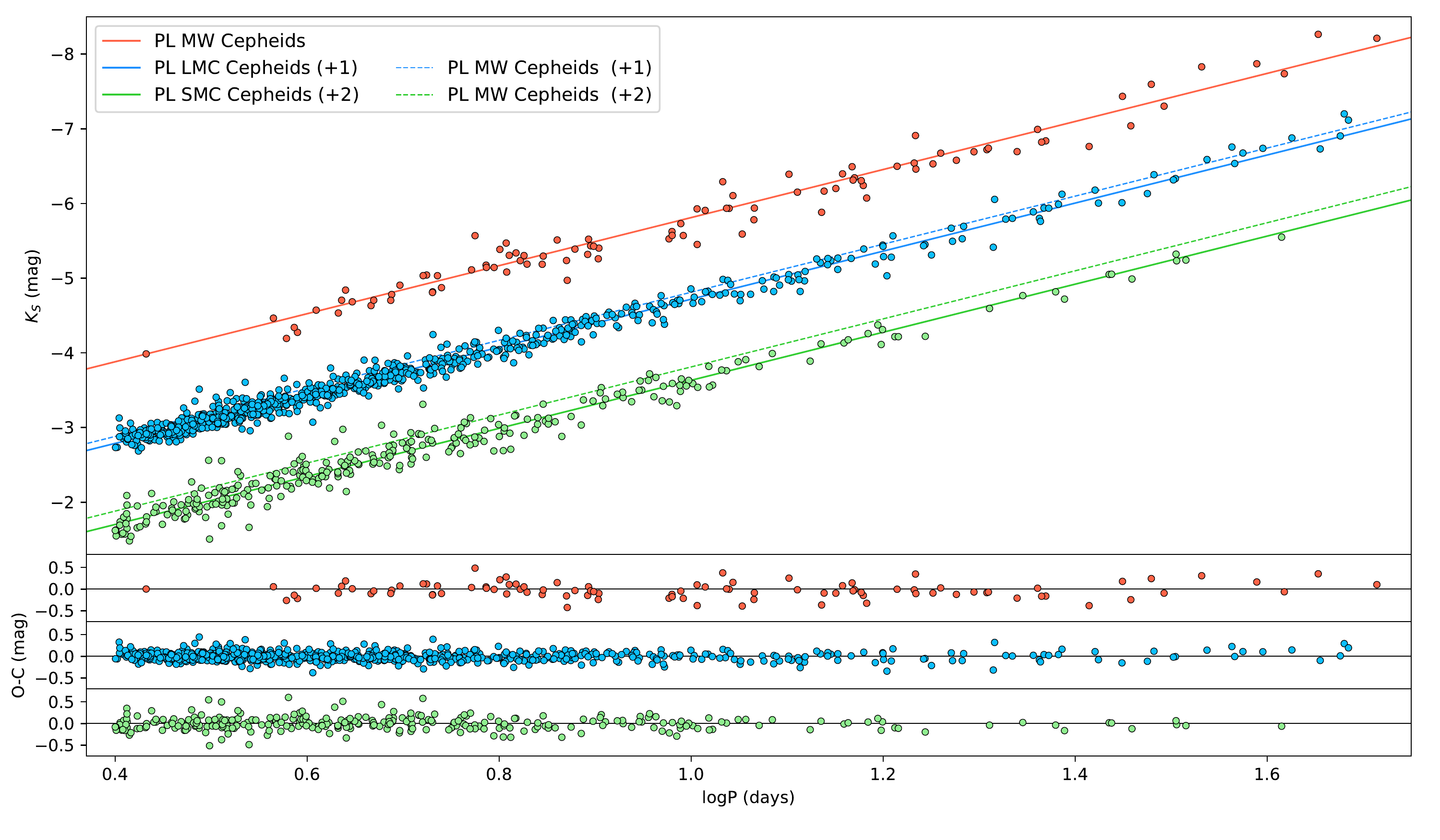}
\caption{Fit of the PL relation in $K_S$ for MW, LMC and SMC Cepheids. The lower panel shows the residual between the Cepheid absolute magnitudes and the corresponding PL fit for each of the three galaxies. The LMC and SMC relations were offset by +1 and +2 mag, respectively, for visualization purposes.}
\label{fig:PL_K}
\end{figure*}

\newpage
\section{The metallicity effect from Milky Way and Magellanic Cloud Cepheids}
\label{sect:gamma_3D}

In this section, we aim at estimating the metallicity term $\gamma$ of the Leavitt law. In Sect. \ref{sect:PL_2D}, we start by fitting the $\alpha$ and $\beta$ coefficients of the PL relation in each of the three galaxies, without considering the metallicity term. In Sect. \ref{sect:PLZ_3D}, we include the metallicity for each galaxy and derive the third term of the PLZ relation by combining the three galaxies. \\

\subsection{The Period-Luminosity relation}
\label{sect:PL_2D}

We adopt the Cepheid samples described in Sect. \ref{sect:sample}. In a first place, we correct apparent magnitudes for the extinction by adopting the reddening law from \citet{Cardelli1989} and \citet{ODonnell1994} assuming $R_V=3.135$ which yields $A_{\lambda} = R_{\lambda} \, E(B-V)$ with $R_I=1.894$, $R_J=0.892$, $R_H=0.553$ and $R_{K_S}=0.363$. We also derive optical and NIR Wesenheit indices \citep{Madore1982} as defined by $W_{VI} = I - 1.526~(V-I)$ and $W_{JK} = K_S - 0.686~(J-K_S)$. Wesenheit magnitudes are particularly convenient for calibrating the PL relation since they are independent of reddening.

\begin{figure*}[]
\centering
\includegraphics[height=4.4cm]{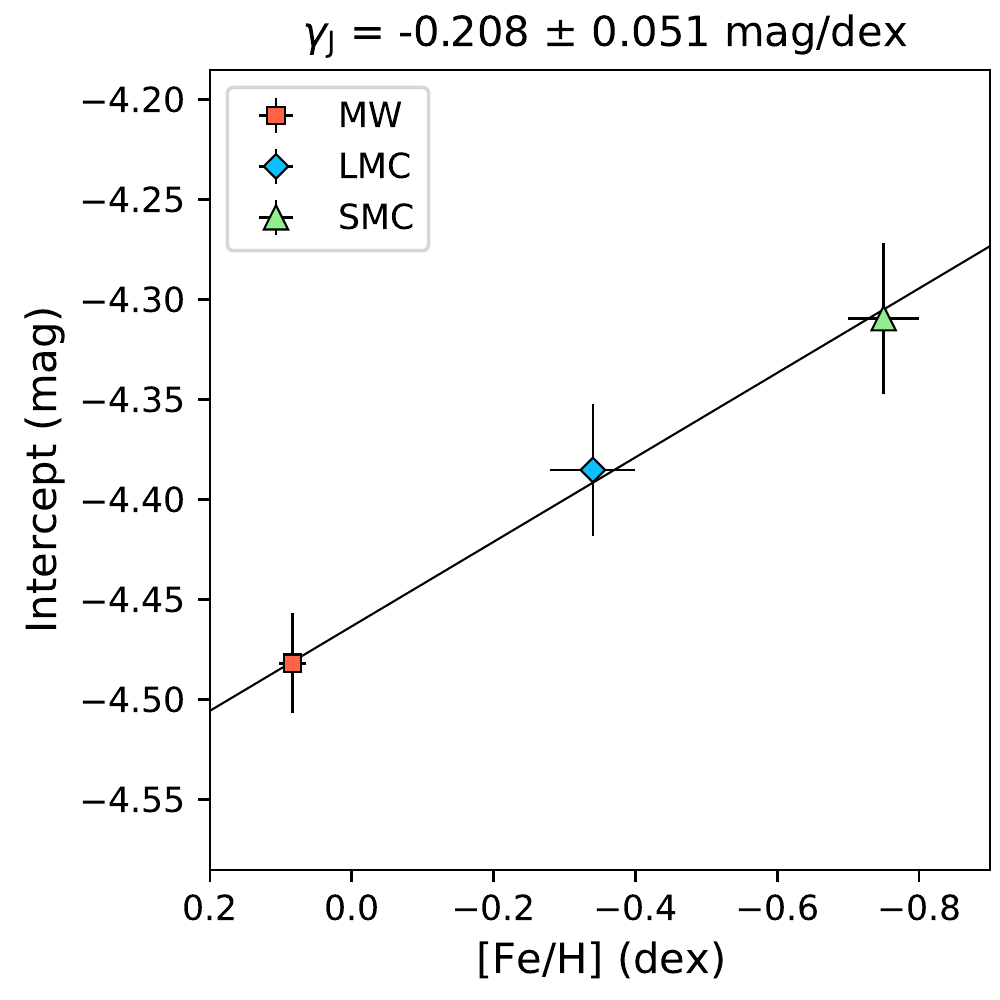}
\includegraphics[height=4.4cm]{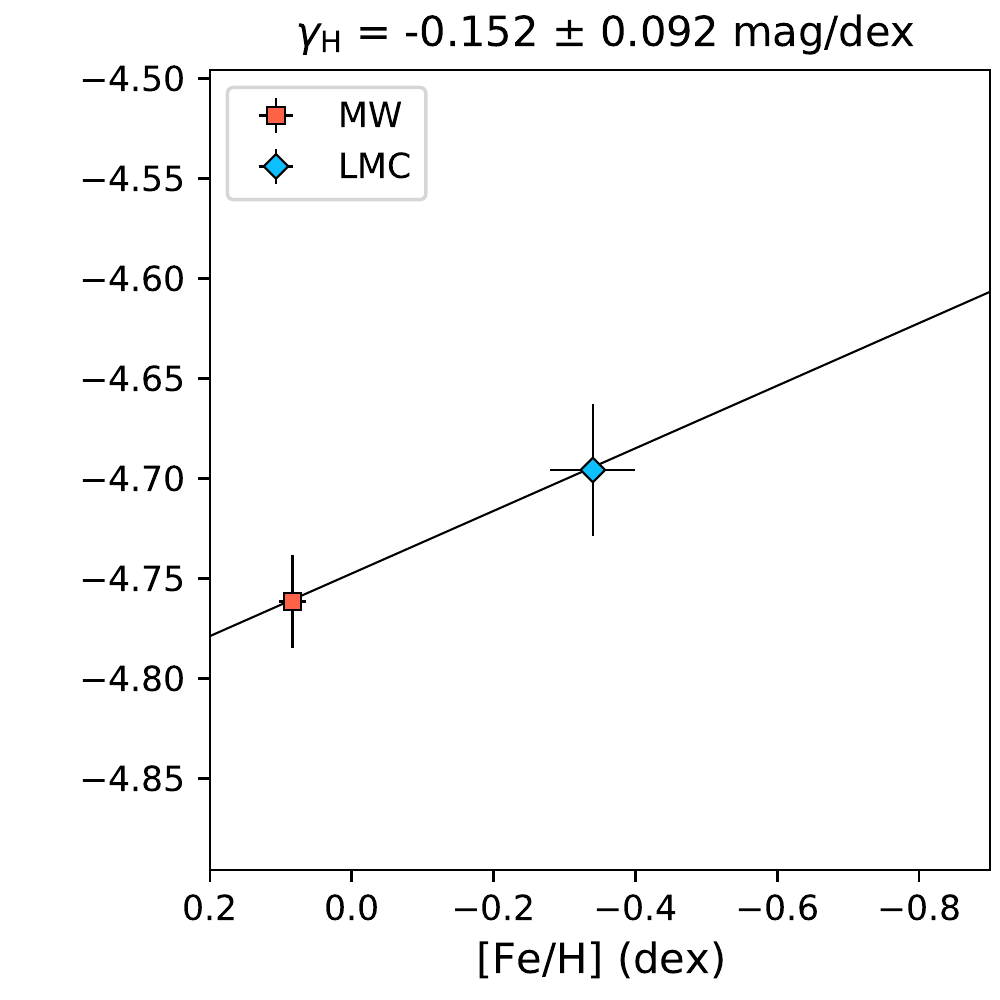}
\includegraphics[height=4.4cm]{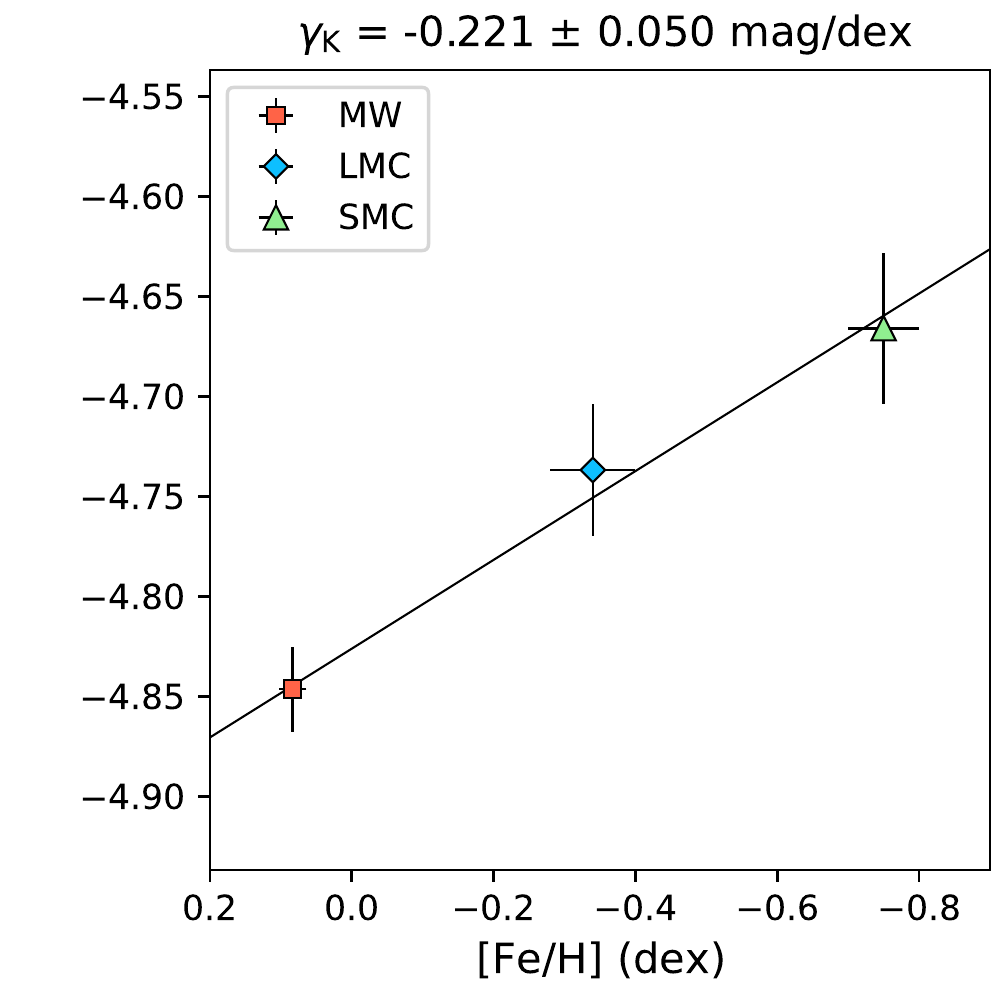}
\includegraphics[height=4.4cm]{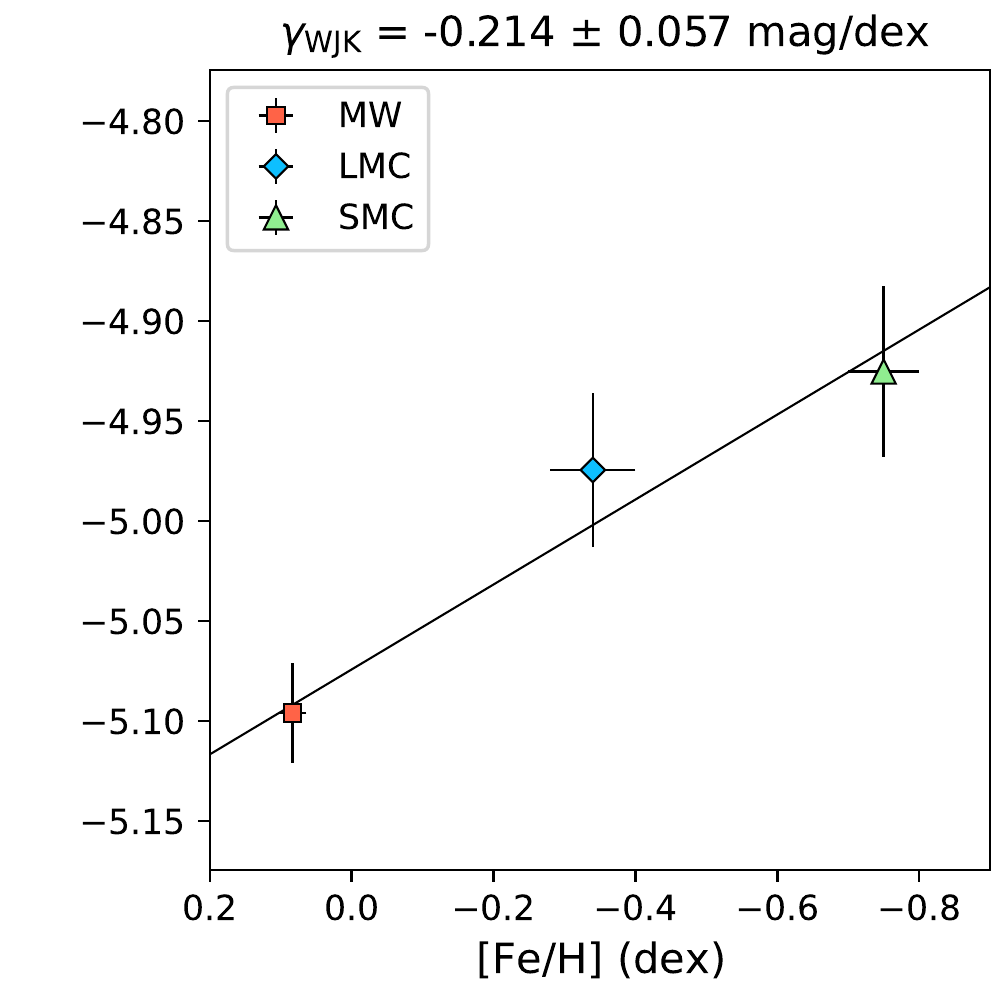}
\includegraphics[height=4.4cm]{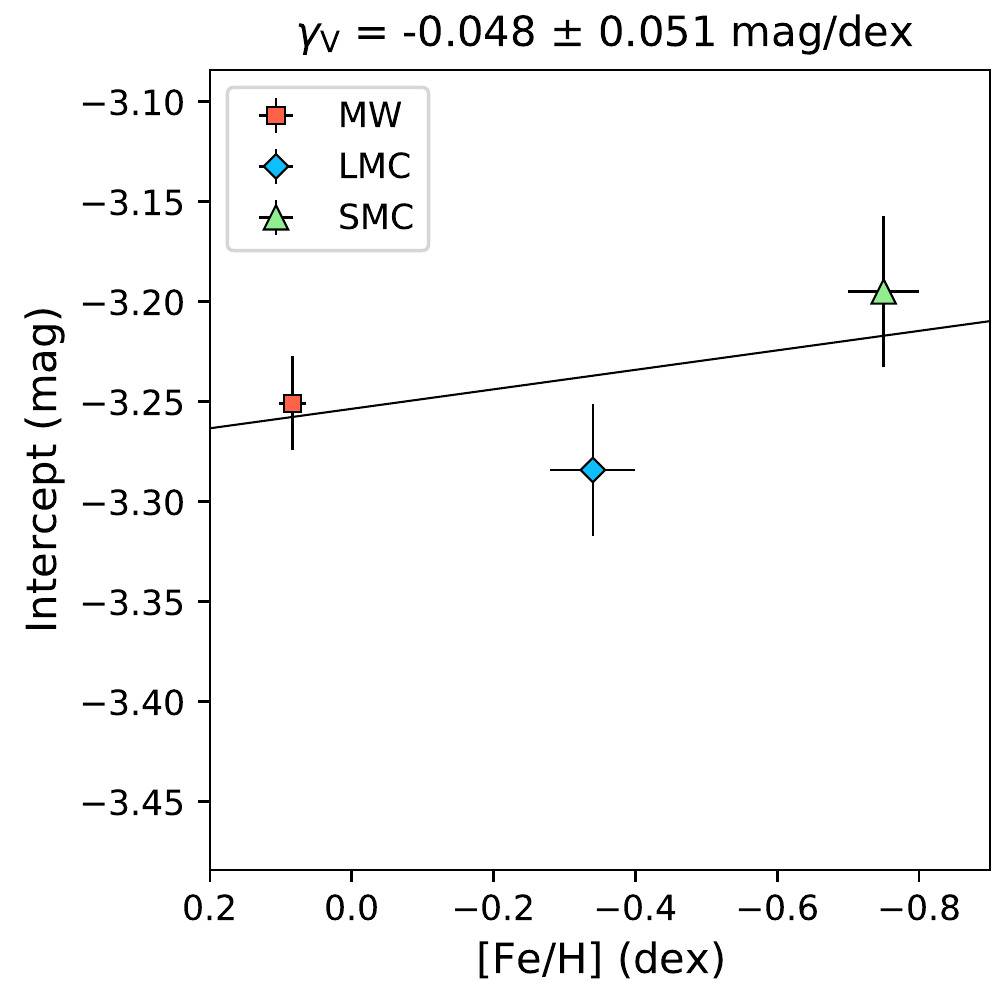}
\includegraphics[height=4.4cm]{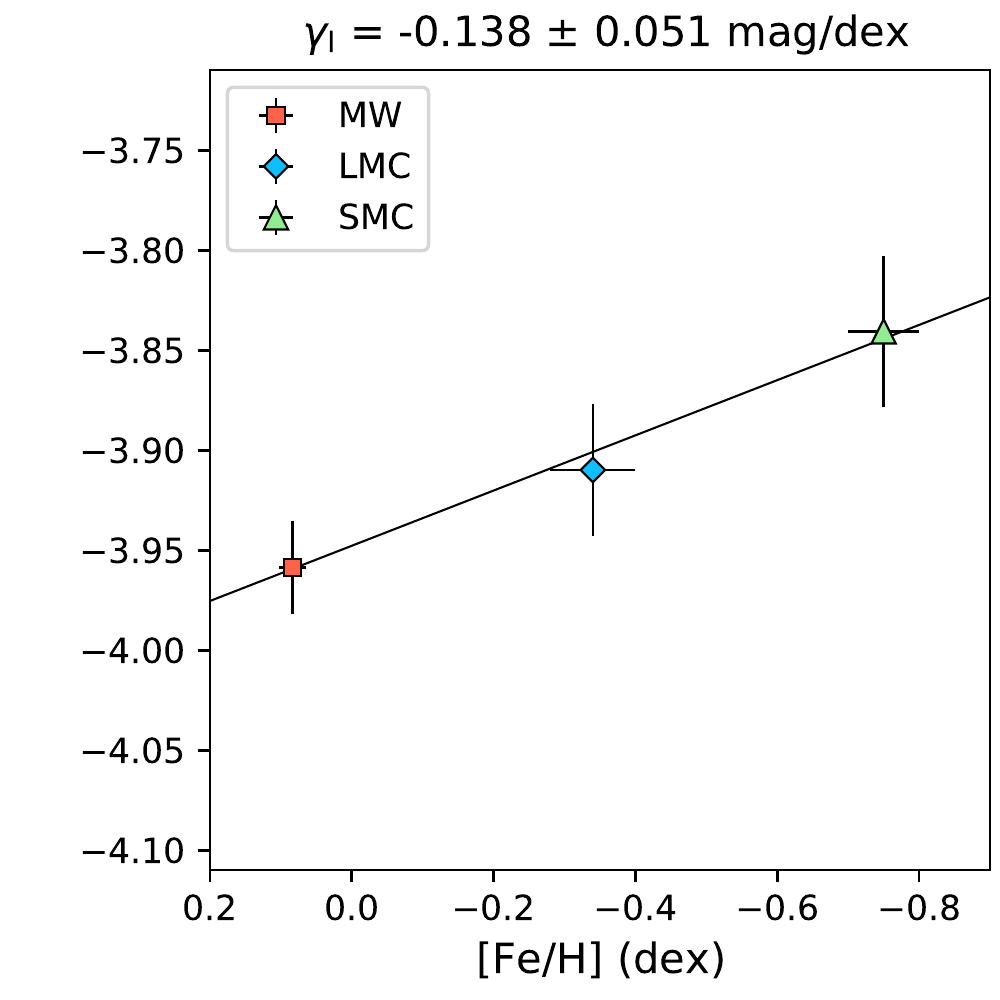}
\includegraphics[height=4.4cm]{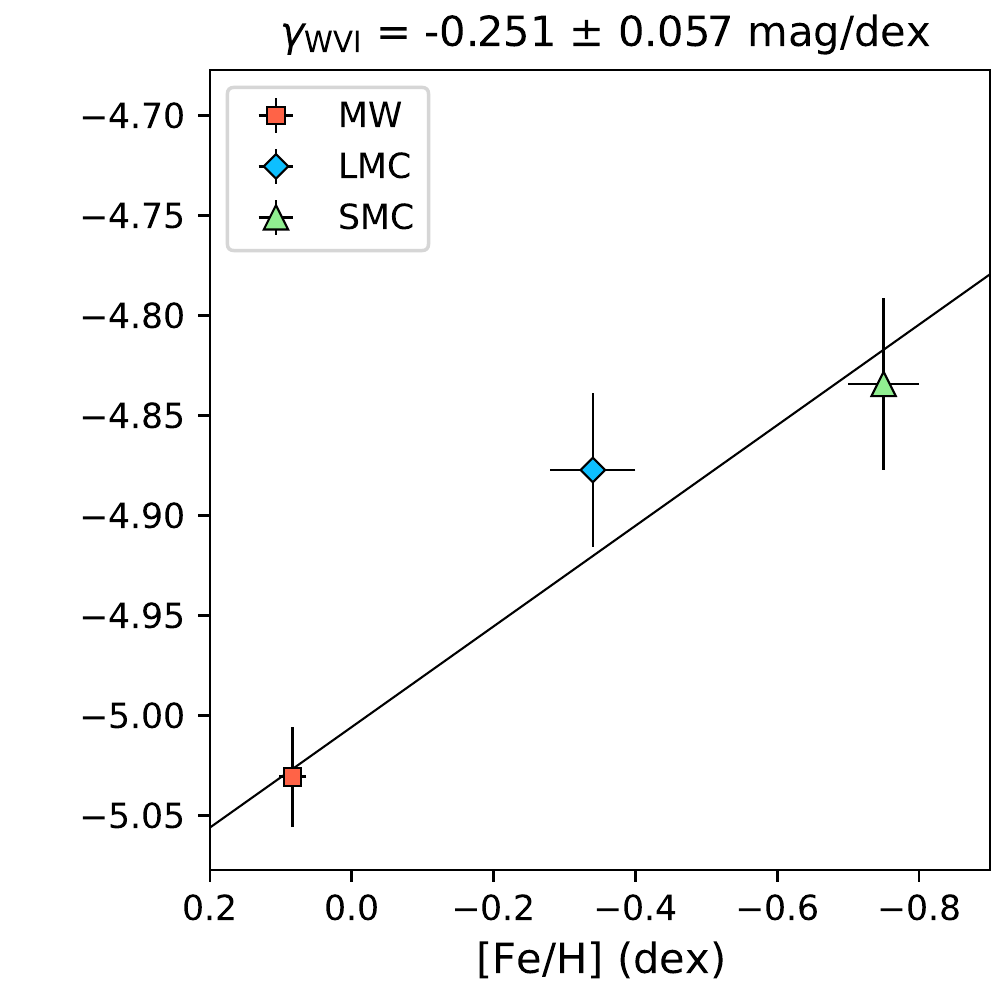}
\caption{Intercept of the PL relation represented as a function of metallicity in $J$, $H$, $K$, $V$, $I$, and Wesenheit bands.}
\label{fig:figA}
\end{figure*}

We account for the width of the instability strip by adding quadratically to the photometry uncertainties the intrinsic scatter in each band: this quantity is obtained by subtracting quadratically the measurement errors (photometric inhomogeneities, differential extinction, geometrical effects, phase corrections, etc) from the scatter of the PL relation: we adopt a width of the instability strip of 0.07 mag in NIR bands ($J$, $H$, $K_S$ and $W_{JK}$) from \citet{Persson2004}, 0.15 mag in $V$ and 0.09 mag in $I$ from \citet{Macri2006} and finally 0.08 mag in $W_{VI}$ from \citet{Madore2017}. We derive the absolute magnitude $M_{\lambda}$ of each Cepheid from their distance $d$ (in kpc) and dereddened apparent magnitude $m_{\lambda}$:
\begin{equation}
M_{\lambda} = m_{\lambda} - 5 \log d -10
\end{equation}

In the Milky Way, the distance is obtained at the first order by taking the inverse of the parallax. In order to avoid biases due to this inversion, we adopt the approach introduced by \citet{Feast1997} and \citet{Arenou1999}, consisting in fitting the Astrometric Based Luminosity (ABL) function instead of absolute magnitudes:
\begin{equation}
\rm ABL = \pi_{\rm(mas)} \, 10^{\, 0.2 m_{\lambda} -2} = 10^{\, M_{\lambda} /5}
\end{equation}
where:
\begin{equation}
M_{\lambda} = \alpha_{\lambda} (\log P - \log P_0) + \beta_{\lambda} 
\end{equation}
We adopt a pivot period of $\log P_0 = 0.7$ which represents the median period of our Cepheid sample. This approach ensures minimum correlations between the fitted coefficients. We perform a 3$\sigma$ clipping procedure on the PL relation to remove possible outliers. 

A non-linearity in the SMC PL relation was highlighted at the short-periods end ($\log P < 0.4$) \citep{Eros1999}. For LMC and SMC Cepheids, \citet{Chown2021} detect a break in the PL relation at $\log P = 0.29$ and also at very long periods ($\log P = 1.72$). Cepheids beyond these limits are found to deviate from the global PL fit and can affect both the slope and the zero-point. Additionally, the short-period edge of the PL relation is potentially affected by first-overtone contamination. In the following, we exclude all Cepheids with periods shorter than 2.5 days ($\log P = 0.4$) and longer than 52 days ($\log P = 1.72$). Finally, we include the systematics on the LMC and SMC distance moduli (respectively 0.026 mag and 0.032 mag) and the photometric zero-points provided in Sect. \ref{sect:sample} on the intercept error. We use the \texttt{curve\_fit} function from the \texttt{scipy} Python library in a Monte Carlo algorithm to derive the PL coefficients and the 16th and 84th percentiles of the distribution to derive the uncertainties. The PL relations derived for each galaxy are provided in Table \ref{table:PL}, where both the slope and intercept are fitted.

In each band, the intercept increases with decreasing metallicity, i.e. it becomes less negative from the MW to the LMC and in turn to the SMC. In the NIR, the intercept changes by $\sim 0.18 \, \rm mag$ between the MW and the SMC, possibly indicating a strong dependence with metallicity. We note that our $K_S$ band calibration in the MW is in good agreement with the result by \citet{Breuval2020} based on \textit{Gaia} DR2 parallaxes. The fit of the PL relation in the $K_S$ band performed in each of the three galaxies is represented in Fig. \ref{fig:PL_K}. \\

\subsection{The Period-Luminosity-Metallicity relation}
\label{sect:PLZ_3D}

\begin{figure}[]
\centering
\includegraphics[height=5.7cm]{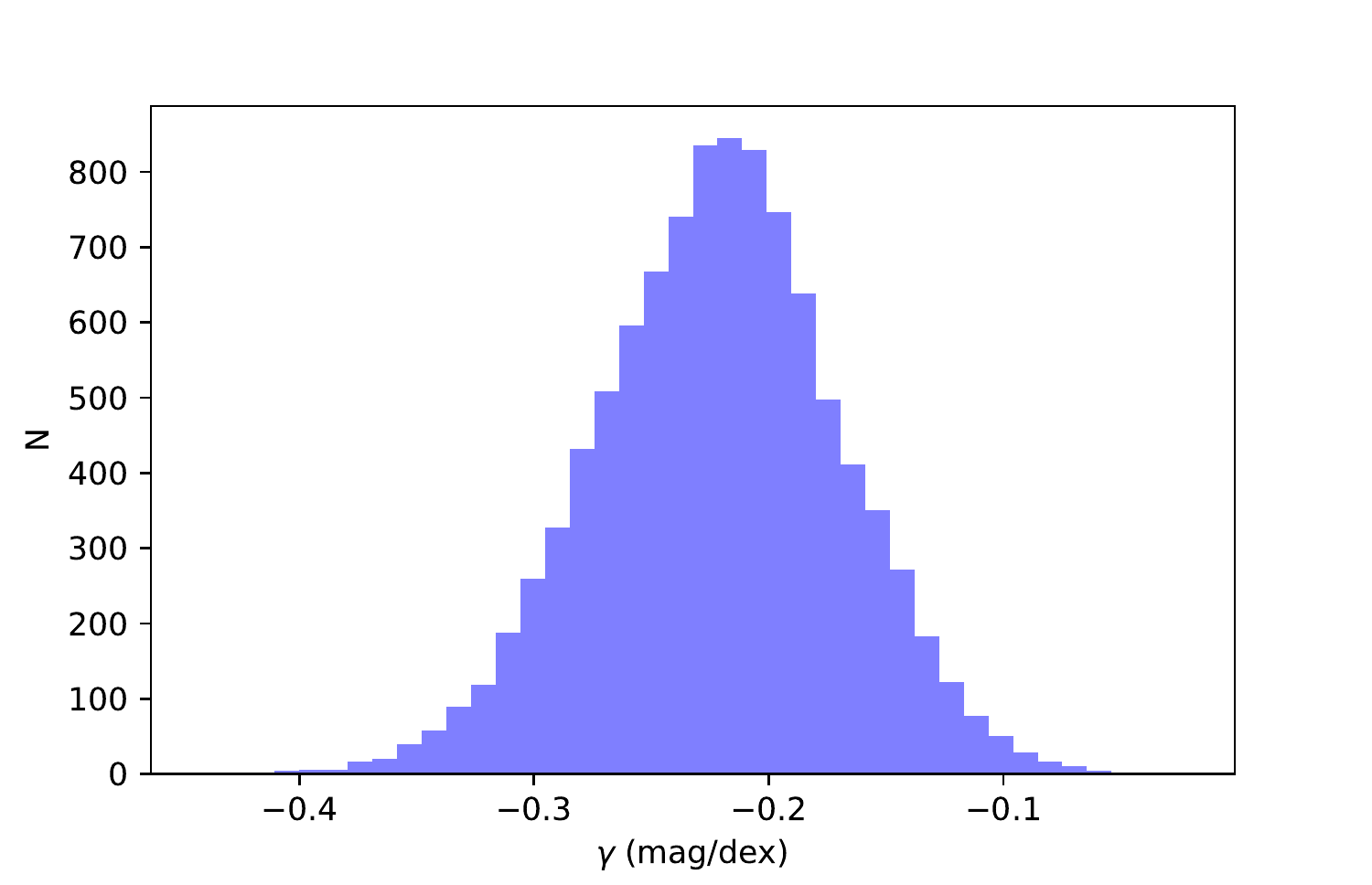}
\caption{Histogram of the $\gamma$ values obtained in the $K_S$ band by the Monte Carlo algorithm iterated 10 000 times.}
\label{fig:histo_gamma}
\end{figure}

In this section, we now calibrate the dependence of the PL intercept $\beta$ with metallicity. First, we fit the PL relation of the form $M = \alpha (\log P - 0.7) + \beta$ in each of the three galaxies separately with a common slope fixed to the LMC value. As in previous section, the systematics due to the LMC and SMC distance and to the photometric zero-point are included in quadrature to the intercept random error. The intercept $\beta$ contains the metallicity term such that:
\begin{equation}
\rm \beta = \gamma\, \rm [Fe/H] + \delta
\label{equat_met}
\end{equation}
In Fig. \ref{fig:figA} are represented the intercepts of the PL relations in the MW, LMC and SMC as a function of metallicity. We fit Eq. \ref{equat_met} with a Monte Carlo algorithm to derive the $\gamma$ and $\delta$ coefficients, and we adopt the 16th and 84th percentiles of the distribution to derive the random errors. A histogram representing the distribution of the $\gamma$ values obtained with the Monte Carlo algorithm is represented in Fig. \ref{fig:histo_gamma}.

\begin{figure}[]
\centering
\includegraphics[height=6.1cm]{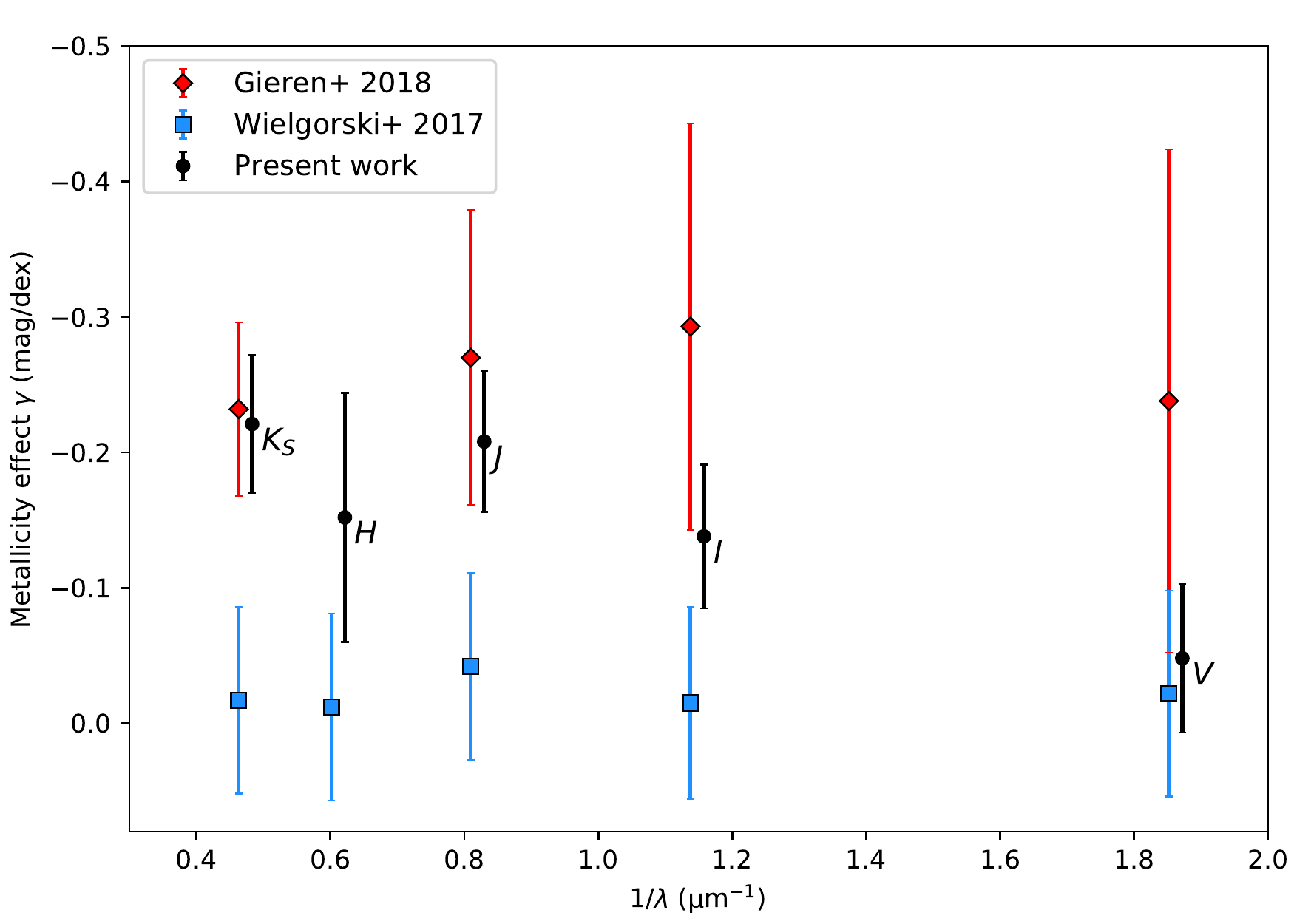}
\caption{Metallicity effect as a function of wavelength, compared with values from the literature. The error bars include the systematics discussed in Sect. \ref{sect:influence_SMC}. \new{For visualization purposes, the X axis was slightly shifted for our values so that the error bars do not overlap, but they correspond to the same wavelength as the literature values.}}
\label{fig:metallicity_wavelength}
\end{figure}

The results of the fit are listed 
in Table \ref{table:3coefs}. In the NIR, we report a strong metallicity effect of $-0.208 \pm 0.051$ mag/dex in $J$, $-0.152 \pm 0.092$ mag/dex in $H$ and $-0.221 \pm 0.050$ mag/dex in $K_S$. The NIR Wesenheit index $W_{JK}$ shows a similar dependence with $-0.214 \pm 0.057$ mag/dex. These results agree by $1 \, \sigma$ with \citet{Gieren2018}, who used the Infrared Surface Brightness Technique \citep{Fouque1997, Storm2011a} to derive the distances to the Cepheids in their MW, LMC and SMC samples, an approach different and independent from the one used in the present study. In the NIR Wesenheit index $W_H$, \citet{Riess2019} find an effect of $-0.170 \pm 0.060$ mag/dex, which is also close to our results in the NIR. In optical bands, we derive a weaker effect than in the NIR with $-0.048 \pm 0.051$ mag/dex in $V$ and of $-0.138 \pm 0.051$ mag/dex in $I$. These values also agree at $1 \, \sigma$ with \citet{Gieren2018}, and the value in $V$ is also consistent at $1 \, \sigma$ with the differential study of LMC and SMC PL relations by \citet{Wielgorski2017}. On average, our results are located between the values by \citet{Wielgorski2017}, consistent with a null metallicity effect, and the work by \citet{Gieren2018} that derive a strong negative effect. In the $H$ band, we derive a metallicity effect weaker than in other NIR bands, likely because it is derived from the MW and LMC samples only (due to the lack of $H$-band photometry for SMC Cepheids). We conclude with the general trend being that the sensitivity to metallicity increases in absolute sense and becomes more negative from optical to NIR wavelengths. This trend is particularly visible in Fig. \ref{fig:metallicity_wavelength}. 

\new{We note that the PL slope was fixed to the LMC value because this sample contains significantly more stars than the two other ones. However, if the slope is fixed to the value found in the Milky Way or in the SMC, the intercepts agree by 0.2\% in NIR and by 1.4\% in optical. Similarly, the $\gamma$ values agree at 0.2 $\sigma$ and 0.8 $\sigma$ in NIR and optical respectively.} \\

\section{Discussion}
\label{sect:discussion}

The metallicity term of the PL relation can be sensitive to many different effects. In this section we study the stability $\gamma$ after varying some parameters. \\

\begin{table}
\flushleft
\caption{Final results of the PLZ fit of the form $M = \alpha (\log P - 0.7) + \delta + \gamma \, \rm [Fe/H]$ and associated uncertainties.}
\begin{tabular}{l | c c | c c | c c}
\hline
\hline
Band 	& $\alpha$ & $\sigma$ & $\delta$ & $\sigma$ & $\gamma$ & $\sigma$	\\
\hline
$V$		& -2.704 & 0.007 & -3.252 & 0.020 & -0.048 & 0.055    \\ 
$I$		& -2.916 & 0.005 & -3.948 & 0.020 & -0.138 & 0.053    \\
$W_{VI}$	& -3.281 & 0.008 & -5.005 & 0.022 & -0.251 & 0.057    \\
$J$		& -3.127 & 0.005 & -4.463 & 0.022 & -0.208 & 0.052    \\
$H$		& -3.160 & 0.005 & -4.748 & 0.020 & -0.152 & 0.092    \\
$K_S$	& -3.217 & 0.004 & -4.826 & 0.019 & -0.221 & 0.051    \\
$W_{JK}$	& -3.273 & 0.008 & -5.075 & 0.022 & -0.214 & 0.057    \\
\hline
\end{tabular}
\flushleft
\tablecomments{The uncertainties include the systematics discussed in Sect. \ref{sect:influence_SMC}.}
\label{table:3coefs}
\end{table}


\subsection{Influence of Gaia EDR3 parallax zero-point}
\label{sect:ZP}

In Sect. \ref{sect:EDR3}, we corrected each \textit{Gaia} EDR3 parallax for their individual zero-point by using the Python tool described in \citet{Lindegren2020a}. However, in \citet{Lindegren2020b}, a uniform parallax zero-point of $-17 \, \rm \mu as$ is derived from quasars. The results of the PLZ fit obtained after adopting this uniform zero-point are provided in the second part of Table \ref{table:PLZ} in appendix. They are consistent at the $1 \sigma$ level with the values derived using the individual zero-point, although it gives a slightly more negative metallicity effect in each band. For example, in $K_S$ we obtain $\gamma = -0.271 \pm 0.051 \, \rm mag/dex$ compared with $\gamma = -0.221 \pm 0.050 \, \rm mag/dex$ with individual zero-points. This effect can be explained by the individual zero-points being on average more negative than $-17 \, \mu as$ for our sample of MW Cepheids.

We also investigate whether the individual zero-point correction by \citet{Lindegren2020a} is adapted to the most distant Cepheids: we remove from our sample the Cepheids with a parallax smaller than $0.3 \, \rm mas$ and derive the PL relation in $K_S$ without these stars. Using this PL relation, we compute the expected parallax of the most distant Cepheids and compare it with the \textit{Gaia} EDR3 parallax corrected by the individual zero-point. We find a good agreement between the predicted parallaxes and the values from \textit{Gaia} EDR3 with the \citet{Lindegren2020a} individual correction. From this study, we confirm that the individual zero-point correction from \citet{Lindegren2020a} is adapted to the most distant MW Cepheids of our sample.  \\

\subsection{Influence of the SMC sample}
\label{sect:influence_SMC}

As mentionned in Sect. \ref{subsec:SMC} and \ref{sect:dist_SMC}, the distance to the core region of the SMC is particularly difficult to measure. From their sample of DEBs, \citet{Graczyk2020} derive an uncertainty of about 2\% for the distance to the SMC core region. These DEB systems are unevenly distributed in the central region of the galaxy and their individual distances show a large dispersion around the mean value, ranging from 57 kpc to 67 kpc (see their Fig. 3), which corresponds to $\sim 16 \%$ of the SMC distance. In order to avoid including Cepheids located too far away from the SMC center, we restricted our sample to a region of radius 0.6 degree around the SMC center. With a smaller radius, the contribution of the SMC sample in the PLZ fit becomes smaller than the MW contribution, therefore we consider that the number of retained SMC Cepheids is insufficient. On the other hand, if we assume a radius larger than 0.6 deg around the SMC center, the number of outlier stars increases and the distance of some Cepheids may not correspond to the distance of the SMC core region. In order to test the validity of our hypothesis, we perform the same PLZ fit with a radius of 0.5 and 0.7 deg around the SMC center and report the coefficients in Table \ref{table:PLZ}. 

After extending the SMC sample to a radius of 0.7 degree around the galaxy centrer, we find $\gamma$ values in very good agreement (better than 1$\sigma$) with the values derived in the initial conditions. When the radius is reduced to 0.5 degree, the metallicity effect still agrees at 1$\sigma$ with the initial conditions in all bands. Considering a smaller region around the SMC center results in a slightly stronger (i.e. more negative) metallicity effect. These results highlight the sensitivity of the metallicity effect with respect to the adopted SMC sample, and in particular to the spatial distribution of the Cepheids considered. Moreover, it emphasizes the necessity to correct each Cepheid distance according to their position in the SMC plane, as we did in Sect. \ref{sect:dist_SMC}. 

We consider the variation of $\gamma$ within a region of $0.5^{\circ} < R < 0.7^{\circ}$ around the SMC center as an additional source of systematic uncertainties: this source of error is at the level of 0.02 mag/dex in optical bands and of 0.01 mag/dex in NIR (see Table \ref{table:PLZ}). We adopt the same additional source of uncertainty for the intercept $\delta$, although the latter coefficient is particularly stable when the radius around the SMC center is changed. These systematics are included in the results presented in Table \ref{table:3coefs}. \\

\section{Conclusions}\label{sec:conclusion}

We build large samples of Cepheids in the Milky Way and in the Magellanic Clouds and make use of the most recent and precise distances available to estimate the metallicity effect on the Cepheid PL relation. In the $K_S$ band we derive an effect of $\gamma = -0.221 \pm 0.051 $ mag/dex, in agreement with the value found by \citet{Gieren2018} \new{but more precise}. In the $V$ band we derive a weaker effect of $\gamma = -0.048 \pm 0.055$ mag/dex, which is consistent with both \citet{Wielgorski2017} and \citet{Gieren2018} within the error bars. We conclude with a non-zero dependence of Cepheid magnitude with metallicity and we confirm its negative sign: metal-rich Cepheids are brighter than metal-poor ones.

The improved precision reached in this work was made possible thanks to the high quality of \textit{Gaia} EDR3 parallaxes and to the new distances of the two Magellanic Clouds obtained by the Araucaria Project. Combining Milky Way and Magellanic Cloud Cepheids also allows to reach a better precision than previous studies based on Magellanic Clouds only, by the larger range of metallicities they cover. A refined analysis of each light curve ensures the use of accurate mean magnitudes. However, the elongated shape of the SMC in the line of sight remains a source of systematic uncertainty in our study, despite continuous efforts to improve our knowledge of its structure. In this study, we assumed a linear dependence of the PL relation with metallicity, but it might as well be non-linear \citep{Gieren2018}. Additional high resolution spectroscopic metallicity measurements of both Milky Way and Magellanic Cloud Cepheids should be carried out in the future to even better constrain the metallicity effect, particularly in the NIR, in our effort to further reduce the systematic uncertainty on the determination of the Hubble constant from the Cepheid-SN Ia method.

\acknowledgements
\new{We thank the anonymous referee for the very constructive comments that helped us to improve the manuscript.} We are grateful to Lucas Macri for providing the photometric transformations. The research leading to these results has received funding from the European Research Council (ERC) under the European Union's Horizon 2020 research and innovation programme under grant agreement No 695099 (project CepBin). This work has made use of data from the European Space Agency (ESA) mission Gaia (http://www.cosmos.esa.int/gaia), processed by the Gaia Data Processing and Analysis Consortium (DPAC, http://www.cosmos.esa.int/web/gaia/dpac/consortium). Funding for the DPAC has been provided by national institutions, in particular the institutions participating in the Gaia Multilateral Agreement.  The authors acknowledge the support of the French Agence Nationale de la Recherche (ANR), under grant ANR-15-CE31-0012-01 (project UnlockCepheids). W.G. and G. P. gratefully acknowledge financial support for this work from the BASAL Centro de Astrofisica y Tecnologias Afines (CATA) AFB- 170002. W.G. and D. G. acknowledge financial support from the Millenium Institute of Astrophysics (MAS) of the Iniciativa Cientifica Milenio del Ministerio de Economia, Fomento y Turismo de Chile, project IC120009. This research made use of Astropy7, a community-developed core Python package for Astronomy \citep{2018AJ....156..123A}. Support from the Polish National Science Centre grants MAESTRO UMO-2017/26/A/ST9/00446 and from the IdPII 2015 0002 64 and DIR/WK/2018/12 grants of the Polish Ministry of Science and Higher Education is also acknowledged. We used the SIMBAD and VIZIER databases and catalog access tool at the CDS, Strasbourg (France), and NASA's Astrophysics Data System Bibliographic Services.

\newpage
~
\newpage
\bibliography{Breuval_Biblio}{}
\bibliographystyle{aasjournal}

 \appendix

 \begin{figure}[]
\centering
\includegraphics[height=8.5cm]{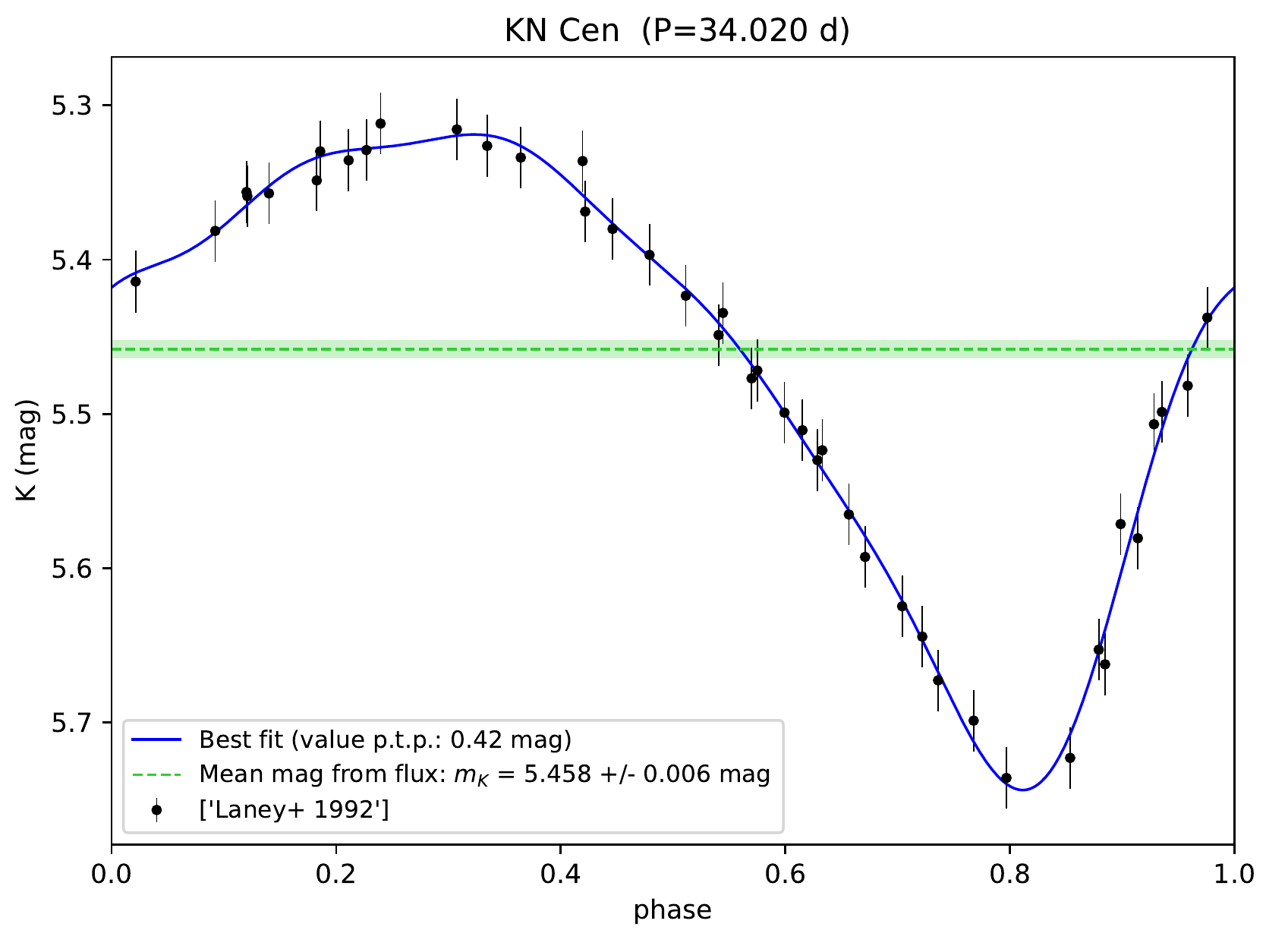}
\caption{Example of a well covered light curve in the $K$ band for the Galactic Cepheid KN Cen. The solid blue line represents the best fit of the light curve, the dashed green line is the mean magnitude derived from the best fit and the green region is the uncertainty on the intensity-averaged mean magnitude.}
\label{fig:good_LC}
\end{figure}

\begin{figure}[]
\centering
\includegraphics[height=8.5cm]{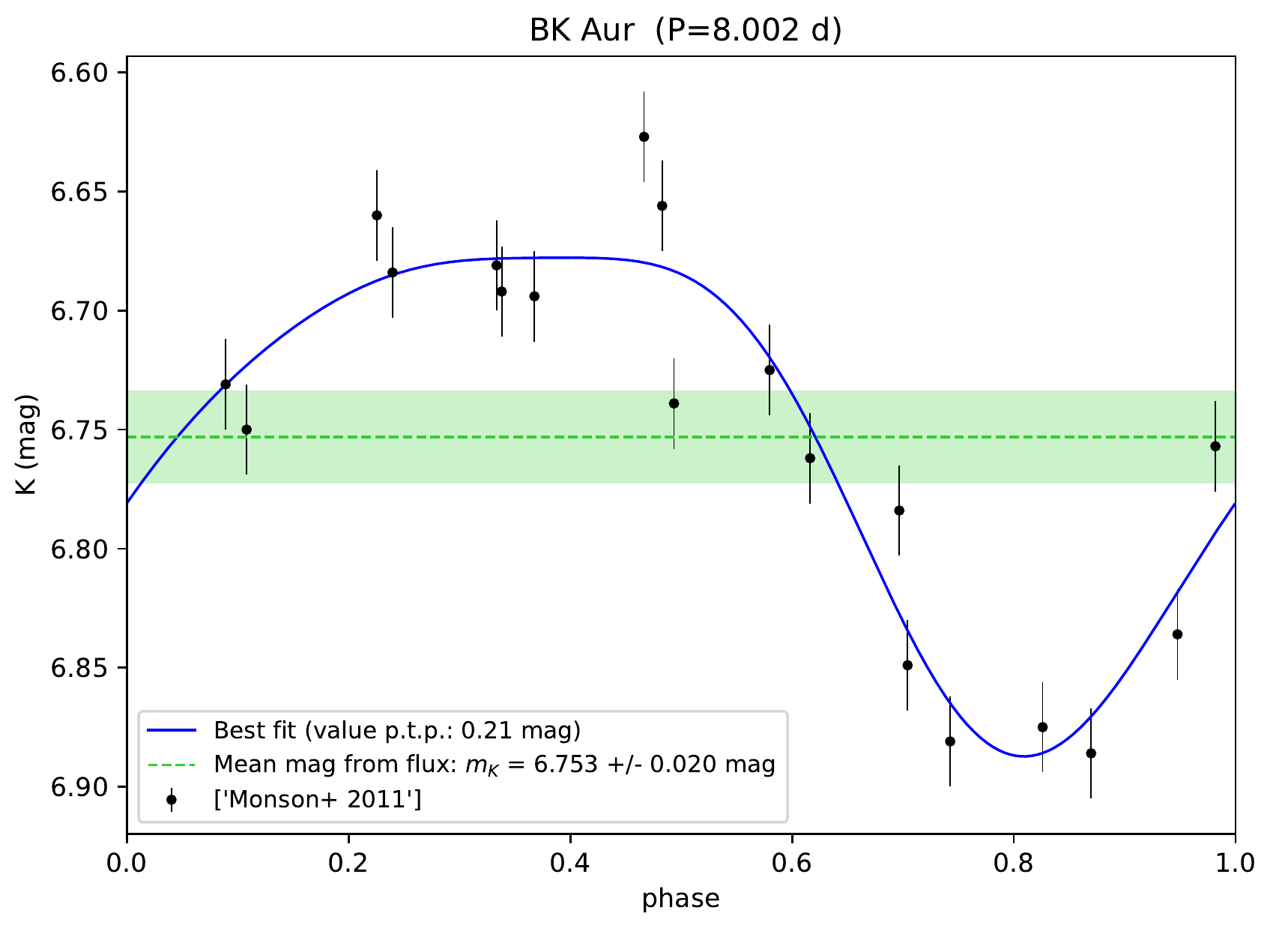}
\caption{Example of a poor-quality light curve in the $K$ band for the Galactic Cepheid BK Aur. The solid blue line represents the best fit of the light curve, the dashed green line is the mean magnitude derived from the best fit and the green region is the uncertainty on the intensity-averaged mean magnitude.}
\label{fig:bad_LC}
\end{figure}

\begin{table*}[]
\footnotesize
\caption{Results of the PLZ linear fit of the form $M = \alpha (\log P - 0.7) + \delta + \gamma \, \rm [Fe/H]$ for Milky Way, Large Magellanic Cloud and Small Magellanic Cloud Cepheids fitted together for different conditions. \new{The slope values are the same as in Table \ref{table:3coefs}.}}
\begin{tabular}{l c c c l}
\hline
\hline
Band 	& $\delta$ 				& $\gamma$			& $N$ & Comments	 \\
\hline
$V$  		& $-3.252 \pm 0.020$ & $-0.048 \pm 0.051$ & 1908 & Initial conditions $^{(\ast)}$ \\ 
$I$   		& $-3.948 \pm 0.020$ & $-0.138 \pm 0.051$ & 1907 & Initial conditions  \\ 
 $W_{VI}$	& $-5.005 \pm 0.022$ & $-0.251 \pm 0.057$ & 1864 & Initial conditions  \\ 
 $J$  	& $-4.463 \pm 0.022$ & $-0.208 \pm 0.051$ & 1196 & Initial conditions   \\ 
 $H$  	& $-4.748 \pm 0.020$ & $-0.152 \pm 0.092$ & 905   & Initial conditions  \\ 
 $K_S$  	& $-4.826 \pm 0.019$ & $-0.221 \pm 0.050$ & 1199 & Initial conditions  \\ 
 $W_{JK}$& $-5.075 \pm 0.022$ & $-0.214 \pm 0.057$ & 1198 & Initial conditions  \\ 
\hline
$V$  		& $-3.274 \pm 0.020$ & $-0.084 \pm 0.051$ & 1908 & Gaia EDR3 parallax ZP = -17$\mu \rm as$ \\ 
$I$   		& $-3.966 \pm 0.020$ & $-0.165 \pm 0.050$ & 1907 & Gaia EDR3 parallax ZP = -17$\mu \rm as$  \\ 
 $W_{VI}$	& $-5.020 \pm 0.022$ & $-0.275 \pm 0.058$ & 1864 & Gaia EDR3 parallax ZP = -17$\mu \rm as$  \\ 
 $J$  	& $-4.495 \pm 0.022$ & $-0.258 \pm 0.052$ & 1196 & Gaia EDR3 parallax ZP = -17$\mu \rm as$   \\ 
 $H$  	& $-4.778 \pm 0.020$ & $-0.241 \pm 0.099$ & 905   & Gaia EDR3 parallax ZP = -17$\mu \rm as$  \\ 
 $K_S$  	& $-4.857 \pm 0.019$ & $-0.271 \pm 0.051$ & 1199 & Gaia EDR3 parallax ZP = -17$\mu \rm as$  \\ 
 $W_{JK}$& $-5.106 \pm 0.022$ & $-0.263 \pm 0.058$ & 1198 & Gaia EDR3 parallax ZP = -17$\mu \rm as$  \\ 
 \hline
$V$  		& $-3.252 \pm 0.020$ & $-0.036 \pm 0.052$ & 1952 & R=0.7$^{\circ}$ around SMC center \\ 
$I$   		& $-3.948 \pm 0.020$ & $-0.130 \pm 0.052$ & 1951 & R=0.7$^{\circ}$ around SMC center  \\ 
 $W_{VI}$	& $-5.005 \pm 0.022$ & $-0.252 \pm 0.058$ & 1908 & R=0.7$^{\circ}$ around SMC center  \\ 
 $J$  	& $-4.464 \pm 0.022$ & $-0.206 \pm 0.052$ & 1241 & R=0.7$^{\circ}$ around SMC center   \\ 
 $K_S$  	& $-4.826 \pm 0.019$ & $-0.218 \pm 0.051$ & 1244 & R=0.7$^{\circ}$ around SMC center  \\ 
 $W_{JK}$& $-5.075 \pm 0.022$ & $-0.208 \pm 0.058$ & 1242 & R=0.7$^{\circ}$ around SMC center  \\ 
 \hline
$V$  		& $-3.250 \pm 0.020$ & $-0.077 \pm 0.051$ & 1845 & R=0.5$^{\circ}$ around SMC center \\ 
$I$   		& $-3.947 \pm 0.020$ & $-0.160 \pm 0.053$ & 1842 & R=0.5$^{\circ}$ around SMC center  \\ 
 $W_{VI}$	& $-5.005 \pm 0.022$ & $-0.251 \pm 0.059$ & 1800 & R=0.5$^{\circ}$ around SMC center  \\ 
 $J$  	& $-4.462 \pm 0.022$ & $-0.225 \pm 0.054$ & 1132 & R=0.5$^{\circ}$ around SMC center   \\ 
 $K_S$  	& $-4.825 \pm 0.019$ & $-0.234 \pm 0.050$ & 1132 & R=0.5$^{\circ}$ around SMC center  \\ 
 $W_{JK}$& $-5.075 \pm 0.022$ & $-0.222 \pm 0.057$ & 1132 & R=0.5$^{\circ}$ around SMC center  \\ 
 \hline
\end{tabular}
\flushleft
\tablecomments{($\ast$) Initial conditions corresponds to \textit{Gaia} EDR3 parallaxes corrected for individual zero-point and SMC Cepheids limited to a radius of 0.6 degree around the SMC center.}
\label{table:PLZ}
\end{table*}

\newpage
~
\newpage

\begin{longtable*}{l c c c c c c c}
\caption{Sample of Milky Way Cepheids and main parameters. Parallaxes from \textit{Gaia} EDR3 include zero point correction. Stars with a RUWE parameter larger than 1.4 were marked with a star and excluded from the sample. \textbf{References}: (F95): reddening from \citet{Fernie1995} multiplied by 0.94; (A12): reddening from \citet{Acharova2012}; (G14): metallicity from \citet{Genovali2014}; (G14b): metallicity from the literature \citep{Genovali2013, Lemasle2007, Luck2011, LuckLambert2011, Pedicelli2010, Romaniello2008, Sziladi2007, Yong2006} rescaled to \citet{Genovali2014} solar abundance; (G15): metallicity from \citet{Genovali2015}.}\\
\hline
\hline
Star	& Period	& $\pi_{\rm EDR3}$	& RUWE	& $E(B-V)$	& Ref.	& [Fe/H] 	& Ref. \\
	&  (day)	&  (mas)			&		&  (mag)		&		&  (dex)	&   \\ 
\hline
\endfirsthead
\multicolumn{8}{c}{\textbf{Table \ref{table:data_MW_parameters}} (continued)} \\
\hline
Star	& Period	& $\pi_{\rm EDR3}$	& RUWE	& $E(B-V)$	& Ref.	& [Fe/H] 	& Ref. \\
	&  (day)	&  (mas)			&		&  (mag)		&		&  (dex)	&   \\ 
\hline 
\endhead
\hline
\endfoot
\endlastfoot
 AA Gem & 11.302 & $0.311 \pm 0.018$ & 1.25 & $0.345 \pm 0.036$ & F95 & $-0.08 \pm 0.05$ & G15 \\
 AC Mon & 8.014 & $0.383 \pm 0.019$ & 1.38 & $0.507 \pm 0.033$ & F95 & $-0.03 \pm 0.06$ & G14b \\
 AD Gem & 3.788 & $0.370 \pm 0.020$ & 0.97 & $0.206 \pm 0.048$ & F95 & $-0.14 \pm 0.06$ & G15 \\
 AD Pup & 13.596 & $0.254 \pm 0.017$ & 1.36 & $0.363 \pm 0.020$ & F95 & $-0.20 \pm 0.15$ & G14b \\
 AE Vel & 7.134 & $0.369 \pm 0.012$ & 0.97 & $0.691 \pm 0.055$ & F95 & $0.14 \pm 0.06$ & G14b \\
 AG Cru & 3.837 & $0.758 \pm 0.020$ & 1.02 & $0.242 \pm 0.020$ & F95 & $0.08 \pm 0.06$ & G14b \\
 AP Pup & 5.084 & $0.924 \pm 0.020$ & 1.05 & $0.250 \pm 0.034$ & F95 & $-0.16 \pm 0.15$ & G14b \\
 AP Sgr & 5.058 & $1.217 \pm 0.024$ & 0.88 & $0.184 \pm 0.015$ & F95 & $0.10 \pm 0.08$ & G14b \\
 AQ Car & 9.769 & $0.361 \pm 0.016$ & 1.07 & $0.168 \pm 0.013$ & F95 & $-0.30 \pm 0.15$ & G14b \\
 AQ Pup & 30.149 & $0.294 \pm 0.023$ & 1.18 & $0.531 \pm 0.017$ & F95 & $0.06 \pm 0.05$ & G15 \\
 AS Per & 4.973 & $0.650 \pm 0.016$ & 1.08 & $0.684 \pm 0.041$ & F95 & $0.14 \pm 0.06$ & G14b \\
 AT Pup & 6.665 & $0.604 \pm 0.016$ & 1.04 & $0.166 \pm 0.011$ & F95 & $-0.22 \pm 0.15$ & G14b \\
 AV Sgr & 15.415 & $0.404 \pm 0.025$ & 0.84 & $1.238 \pm 0.027$ & F95 & $0.35 \pm 0.17$ & G15 \\
 AW Per & 6.464 & $1.093 \pm 0.029$ & 1.16 & $0.479 \pm 0.016$ & F95 & $0.04 \pm 0.06$ & G14b \\
 AY Cas & 2.872 & $0.414 \pm 0.019$ & 1.07 & $0.760 \pm 0.049$ & F95 & $0.02 \pm 0.06$ & G14b \\
 AY Cen & 5.310 & $0.574 \pm 0.014$ & 0.95 & $0.357 \pm 0.066$ & F95 & $0.08 \pm 0.06$ & G14b \\
 AY Sgr & 6.570 & $0.546 \pm 0.019$ & 0.85 & $0.840 \pm 0.009$ & F95 & $0.11 \pm 0.06$ & G15 \\
 BB Her & 7.508 & $0.280 \pm 0.015$ & 1.06 & $0.392 \pm 0.039$ & A12 & $0.26 \pm 0.06$ & G14b \\
 BB Sgr & 6.637 & $1.188 \pm 0.024$ & 0.82 & $0.285 \pm 0.011$ & F95 & $0.08 \pm 0.08$ & G14b \\
 BE Mon & 2.706 & $0.504 \pm 0.017$ & 1.17 & $0.549 \pm 0.036$ & F95 & $0.05 \pm 0.09$ & G15 \\
 BF Oph & 4.068 & $1.189 \pm 0.024$ & 0.84 & $0.261 \pm 0.016$ & F95 & $0.14 \pm 0.06$ & G14b \\
 BG Vel & 6.924 & $1.045 \pm 0.017$ & 0.99 & $0.434 \pm 0.011$ & F95 & $-0.10 \pm 0.15$ & G14b \\
 BK Aur & 8.002 & $0.426 \pm 0.015$ & 1.01 & $0.393 \pm 0.026$ & F95 & $-0.07 \pm 0.15$ & G14b \\
 BM Per & 22.952 & $0.334 \pm 0.022$ & 0.99 & $0.919 \pm 0.059$ & F95 & $0.23 \pm 0.06$ & G14b \\
 BM Pup & 7.199 & $0.302 \pm 0.013$ & 1.18 & $0.575 \pm 0.058$ & F95 & $-0.07 \pm 0.08$ & G15 \\
 BN Pup & 13.673 & $0.301 \pm 0.015$ & 1.25 & $0.422 \pm 0.017$ & F95 & $0.03 \pm 0.05$ & G15 \\
 BP Cas & 6.273 & $0.442 \pm 0.013$ & 1.02 & $0.864 \pm 0.014$ & F95 & $0.09 \pm 0.06$ & G14b \\
 BZ Cyg & 10.142 & $0.500 \pm 0.014$ & 1.14 & $0.832 \pm 0.018$ & F95 & $0.19 \pm 0.08$ & G14b \\
 CD Cas & 7.801 & $0.412 \pm 0.014$ & 1.06 & $0.745 \pm 0.012$ & F95 & $0.13 \pm 0.06$ & G14b \\
 CD Cyg & 17.074 & $0.394 \pm 0.016$ & 1.01 & $0.512 \pm 0.021$ & F95 & $0.15 \pm 0.06$ & G14b \\
 CE Pup & 49.326 & $0.114 \pm 0.014$ & 0.82 & $0.740 \pm 0.074$ & A12 & $-0.04 \pm 0.09$ & G14 \\
 CF Cas & 4.875 & $0.316 \pm 0.012$ & 1.04 & $0.556 \pm 0.021$ & F95 & $0.02 \pm 0.06$ & G14b \\
 CG Cas & 4.366 & $0.296 \pm 0.017$ & 1.03 & $0.667 \pm 0.009$ & F95 & $0.09 \pm 0.06$ & G14b \\
 CK Sct & 7.415 & $0.490 \pm 0.020$ & 0.97 & $0.816 \pm 0.024$ & F95 & $0.21 \pm 0.06$ & G14b \\
 CN Car & 4.933 & $0.342 \pm 0.014$ & 0.96 & $0.438 \pm 0.049$ & F95 & $0.21 \pm 0.06$ & G14b \\
 CP Cep & 17.859 & $0.279 \pm 0.021$ & 1.01 & $0.681 \pm 0.045$ & F95 & $-0.01 \pm 0.08$ & G14b \\
 CR Cep & 6.233 & $0.699 \pm 0.013$ & 1.06 & $0.704 \pm 0.009$ & F95 & $-0.06 \pm 0.08$ & G14b \\
 CR Ser & 5.301 & $0.578 \pm 0.020$ & 1.19 & $0.974 \pm 0.017$ & F95 & $0.12 \pm 0.08$ & G15 \\
 CS Mon & 6.732 & $0.324 \pm 0.014$ & 1.04 & $0.528 \pm 0.032$ & F95 & $-0.08 \pm 0.06$ & G14b \\
 CS Ori & 3.889 & $0.257 \pm 0.022$ & 1.33 & $0.373 \pm 0.030$ & F95 & $-0.25 \pm 0.06$ & G15 \\
 CS Vel & 5.905 & $0.272 \pm 0.016$ & 0.91 & $0.716 \pm 0.027$ & F95 & $0.12 \pm 0.06$ & G14b \\
 CV Mon & 5.379 & $0.601 \pm 0.015$ & 1.10 & $0.705 \pm 0.018$ & F95 & $0.09 \pm 0.09$ & G15 \\
 CY Car & 4.266 & $0.427 \pm 0.011$ & 0.93 & $0.409 \pm 0.043$ & F95 & $0.11 \pm 0.06$ & G14b \\
 CY Cas & 14.377 & $0.255 \pm 0.019$ & 1.07 & $0.952 \pm 0.008$ & F95 & $0.06 \pm 0.08$ & G14b \\
 CZ Cas & 5.664 & $0.292 \pm 0.016$ & 0.96 & $0.761 \pm 0.030$ & F95 & $0.07 \pm 0.06$ & G14b \\
 DD Cas & 9.812 & $0.346 \pm 0.013$ & 1.05 & $0.486 \pm 0.016$ & F95 & $0.10 \pm 0.08$ & G14b \\
 DF Cas & 3.832 & $0.374 \pm 0.014$ & 1.05 & $0.564 \pm 0.049$ & F95 & $0.13 \pm 0.08$ & G14b \\
 DW Per & 3.650 & $0.296 \pm 0.019$ & 1.31 & $0.620 \pm 0.033$ & F95 & $-0.05 \pm 0.06$ & G14b \\
 EK Mon & 3.958 & $0.376 \pm 0.021$ & 1.16 & $0.547 \pm 0.003$ & F95 & $-0.05 \pm 0.15$ & G14b \\
 ER Car & 7.719 & $0.869 \pm 0.015$ & 0.82 & $0.111 \pm 0.016$ & F95 & $0.15 \pm 0.06$ & G14b \\
 EX Vel & 13.234 & $0.204 \pm 0.015$ & 0.94 & $0.728 \pm 0.052$ & F95 & $0.07 \pm 0.06$ & G14b \\
 FI Car & 13.458 & $0.242 \pm 0.019$ & 0.99 & $0.694 \pm 0.007$ & F95 & $0.31 \pm 0.06$ & G14b \\
 FM Aql & 6.114 & $1.014 \pm 0.026$ & 1.26 & $0.635 \pm 0.019$ & F95 & $0.24 \pm 0.06$ & G14b \\
 FN Aql & 9.482 & $0.736 \pm 0.025$ & 1.12 & $0.486 \pm 0.008$ & F95 & $-0.06 \pm 0.06$ & G14b \\
 GH Cyg & 7.818 & $0.417 \pm 0.014$ & 1.07 & $0.608 \pm 0.023$ & F95 & $0.21 \pm 0.06$ & G14b \\
 GI Cyg & 5.783 & $0.273 \pm 0.017$ & 1.01 & $0.734 \pm 0.073$ & F95 & $0.27 \pm 0.06$ & G14b \\
 GQ Ori & 8.616 & $0.408 \pm 0.021$ & 0.87 & $0.224 \pm 0.013$ & F95 & $0.20 \pm 0.08$ & G15 \\
 GU Nor & 3.453 & $0.565 \pm 0.015$ & 0.87 & $0.683 \pm 0.029$ & F95 & $0.08 \pm 0.06$ & G15 \\
 GX Car & 7.197 & $0.459 \pm 0.013$ & 1.02 & $0.380 \pm 0.008$ & F95 & $0.14 \pm 0.06$ & G14b \\
 GY Sge & 51.790 & $0.342 \pm 0.023$ & 0.95 & $1.183 \pm 0.111$ & F95 & $0.29 \pm 0.06$ & G14b \\
 HW Car & 9.199 & $0.397 \pm 0.012$ & 0.94 & $0.181 \pm 0.018$ & F95 & $0.09 \pm 0.06$ & G14b \\
 IQ Nor & 8.220 & $0.535 \pm 0.018$ & 0.97 & $0.676 \pm 0.044$ & F95 & $0.22 \pm 0.07$ & G15 \\
 IT Car & 7.533 & $0.702 \pm 0.020$ & 1.08 & $0.212 \pm 0.016$ & F95 & $0.14 \pm 0.06$ & G14b \\
 KK Cen & 12.180 & $0.152 \pm 0.018$ & 1.03 & $0.555 \pm 0.033$ & F95 & $0.24 \pm 0.06$ & G14b \\
 KN Cen & 34.020 & $0.251 \pm 0.018$ & 1.03 & $0.728 \pm 0.040$ & F95 & $0.55 \pm 0.12$ & G15 \\
 KQ Sco & 28.705 & $0.472 \pm 0.021$ & 0.91 & $0.852 \pm 0.041$ & F95 & $0.52 \pm 0.08$ & G15 \\
 LS Pup & 14.147 & $0.214 \pm 0.016$ & 1.25 & $0.452 \pm 0.009$ & F95 & $-0.12 \pm 0.11$ & G15 \\
 MW Cyg & 5.955 & $0.542 \pm 0.019$ & 1.21 & $0.651 \pm 0.039$ & F95 & $0.09 \pm 0.08$ & G14b \\
 MZ Cen & 10.354 & $0.221 \pm 0.017$ & 0.84 & $0.782 \pm 0.077$ & F95 & $0.27 \pm 0.10$ & G15 \\
 QY Cen & 17.752 & $0.293 \pm 0.021$ & 1.02 & $1.213 \pm 0.216$ & F95 & $0.24 \pm 0.06$ & G14b \\
 R Cru & 5.826 & $1.078 \pm 0.028$ & 1.16 & $0.156 \pm 0.012$ & F95 & $0.13 \pm 0.06$ & G14b \\
 R Mus & 7.510 & $1.076 \pm 0.018$ & 1.07 & $0.149 \pm 0.030$ & F95 & $-0.08 \pm 0.06$ & G14b \\
 R TrA & 3.389 & $1.560 \pm 0.016$ & 0.89 & $0.167 \pm 0.025$ & F95 & $0.19 \pm 0.06$ & G14b \\
 RR Lac & 6.416 & $0.424 \pm 0.015$ & 1.10 & $0.267 \pm 0.023$ & F95 & $0.04 \pm 0.06$ & G14b \\
 RS Nor & 6.198 & $0.472 \pm 0.017$ & 0.94 & $0.577 \pm 0.036$ & F95 & $0.18 \pm 0.08$ & G15 \\
 RS Ori & 7.567 & $0.589 \pm 0.030$ & 1.12 & $0.332 \pm 0.010$ & F95 & $0.11 \pm 0.09$ & G15 \\
 RS Pup & 41.480 & $0.581 \pm 0.017$ & 1.16 & $0.451 \pm 0.010$ & F95 & $0.07 \pm 0.15$ & G14b \\
 RU Sct & 19.704 & $0.526 \pm 0.024$ & 0.87 & $0.914 \pm 0.017$ & F95 & $0.14 \pm 0.04$ & G15 \\
 RV Sco & 6.061 & $1.257 \pm 0.021$ & 0.81 & $0.343 \pm 0.007$ & F95 & $0.11 \pm 0.06$ & G14b \\
 RW Cas & 14.795 & $0.335 \pm 0.019$ & 1.26 & $0.440 \pm 0.032$ & F95 & $0.22 \pm 0.08$ & G14b \\
 RX Aur & 11.624 & $0.654 \pm 0.021$ & 0.98 & $0.254 \pm 0.020$ & F95 & $0.10 \pm 0.06$ & G14b \\
 RY CMa & 4.678 & $0.825 \pm 0.029$ & 1.29 & $0.238 \pm 0.016$ & F95 & $0.00 \pm 0.15$ & G14b \\
 RY Sco & 20.323 & $0.764 \pm 0.032$ & 0.73 & $0.654 \pm 0.044$ & F95 & $0.01 \pm 0.06$ & G15 \\
 RY Vel & 28.136 & $0.376 \pm 0.021$ & 1.07 & $0.539 \pm 0.012$ & F95 & $-0.05 \pm 0.15$ & G14b \\
 RZ Vel & 20.398 & $0.661 \pm 0.017$ & 1.24 & $0.301 \pm 0.011$ & F95 & $0.05 \pm 0.15$ & G14b \\
 S Cru & 4.690 & $1.342 \pm 0.024$ & 0.94 & $0.172 \pm 0.014$ & F95 & $0.11 \pm 0.06$ & G14b \\
 S Nor & 9.754 & $1.099 \pm 0.022$ & 0.88 & $0.182 \pm 0.008$ & F95 & $0.02 \pm 0.09$ & G14b \\
 S TrA & 6.324 & $1.120 \pm 0.022$ & 1.04 & $0.086 \pm 0.010$ & F95 & $0.21 \pm 0.06$ & G14b \\
 SS CMa & 12.361 & $0.307 \pm 0.013$ & 1.11 & $0.551 \pm 0.012$ & F95 & $0.06 \pm 0.04$ & G15 \\
 SS Sct & 3.671 & $0.934 \pm 0.023$ & 0.84 & $0.340 \pm 0.022$ & F95 & $0.14 \pm 0.06$ & G14b \\
 ST Tau & 4.034 & $0.916 \pm 0.034$ & 1.35 & $0.328 \pm 0.006$ & F95 & $-0.14 \pm 0.15$ & G14b \\
 ST Vel & 5.858 & $0.384 \pm 0.015$ & 1.19 & $0.530 \pm 0.024$ & F95 & $-0.14 \pm 0.15$ & G14b \\
 SV Mon & 15.233 & $0.464 \pm 0.032$ & 1.01 & $0.264 \pm 0.021$ & F95 & $0.12 \pm 0.08$ & G15 \\
 SV Vel & 14.097 & $0.434 \pm 0.018$ & 1.02 & $0.376 \pm 0.024$ & F95 & $0.12 \pm 0.06$ & G14b \\
 SV Vul & 44.993 & $0.402 \pm 0.021$ & 1.20 & $0.474 \pm 0.024$ & F95 & $0.05 \pm 0.08$ & G14b \\
 SW Cas & 5.441 & $0.461 \pm 0.012$ & 1.12 & $0.475 \pm 0.027$ & F95 & $-0.03 \pm 0.08$ & G14b \\
 SW Vel & 23.407 & $0.413 \pm 0.018$ & 1.05 & $0.338 \pm 0.009$ & F95 & $-0.15 \pm 0.15$ & G14b \\
 SX Car & 4.860 & $0.515 \pm 0.022$ & 1.25 & $0.323 \pm 0.026$ & F95 & $0.05 \pm 0.06$ & G14b \\
 SX Per & 4.290 & $0.313 \pm 0.019$ & 1.19 & $0.537 \pm 0.046$ & F95 & $-0.03 \pm 0.06$ & G14b \\
 SX Vel & 9.550 & $0.501 \pm 0.019$ & 1.02 & $0.237 \pm 0.014$ & F95 & $-0.18 \pm 0.15$ & G14b \\
 SY Aur & 10.145 & $0.462 \pm 0.020$ & 1.08 & $0.386 \pm 0.040$ & F95 & $-0.07 \pm 0.15$ & G14b \\
 SZ Aql & 17.141 & $0.525 \pm 0.020$ & 0.94 & $0.553 \pm 0.022$ & F95 & $0.18 \pm 0.08$ & G14b \\
 SZ Cas & 13.639 & $0.407 \pm 0.017$ & 1.01 & $0.713 \pm 0.060$ & F95 & $0.07 \pm 0.06$ & G14b \\
 SZ Cyg & 15.110 & $0.445 \pm 0.012$ & 0.96 & $0.594 \pm 0.004$ & F95 & $0.09 \pm 0.08$ & G14b \\
 T Ant & 5.898 & $0.312 \pm 0.014$ & 1.18 & $0.300 \pm 0.030$ & A12 & $-0.20 \pm 0.06$ & G14b \\
 T Cru & 6.733 & $1.222 \pm 0.014$ & 0.82 & $0.191 \pm 0.022$ & F95 & $0.14 \pm 0.06$ & G14b \\
 T Vel & 4.640 & $0.940 \pm 0.016$ & 0.93 & $0.282 \pm 0.018$ & F95 & $-0.02 \pm 0.15$ & G14b \\
 T Vul & 4.435 & $1.719 \pm 0.058$ & 1.20 & $0.092 \pm 0.017$ & F95 & $0.01 \pm 0.08$ & G14b \\
 TT Aql & 13.755 & $0.997 \pm 0.023$ & 1.08 & $0.487 \pm 0.024$ & F95 & $0.22 \pm 0.06$ & G14b \\
 TV CMa & 4.670 & $0.420 \pm 0.015$ & 1.20 & $0.574 \pm 0.029$ & F95 & $0.01 \pm 0.07$ & G15 \\
 TV Cam & 5.295 & $0.237 \pm 0.018$ & 1.11 & $0.560 \pm 0.023$ & F95 & $0.04 \pm 0.06$ & G14b \\
 TW CMa & 6.995 & $0.384 \pm 0.019$ & 1.15 & $0.374 \pm 0.033$ & F95 & $0.04 \pm 0.09$ & G15 \\
 TW Nor & 10.786 & $0.360 \pm 0.020$ & 0.89 & $1.190 \pm 0.023$ & F95 & $0.27 \pm 0.10$ & G15 \\
 TX Cen & 17.098 & $0.332 \pm 0.018$ & 0.94 & $0.941 \pm 0.038$ & F95 & $0.44 \pm 0.12$ & G15 \\
 TX Cyg & 14.710 & $0.829 \pm 0.019$ & 0.95 & $1.123 \pm 0.005$ & F95 & $0.20 \pm 0.08$ & G14b \\
 TY Sct & 11.053 & $0.371 \pm 0.016$ & 0.91 & $0.930 \pm 0.017$ & F95 & $0.37 \pm 0.06$ & G14b \\
 TZ Mon & 7.428 & $0.298 \pm 0.015$ & 1.24 & $0.434 \pm 0.023$ & F95 & $-0.02 \pm 0.07$ & G15 \\
 TZ Mus & 4.945 & $0.266 \pm 0.020$ & 1.00 & $0.676 \pm 0.020$ & F95 & $0.10 \pm 0.06$ & G14b \\
 U Car & 38.829 & $0.561 \pm 0.023$ & 1.23 & $0.276 \pm 0.013$ & F95 & $0.17 \pm 0.09$ & G14b \\
 U Nor & 12.644 & $0.625 \pm 0.019$ & 0.98 & $0.868 \pm 0.038$ & F95 & $0.07 \pm 0.09$ & G14b \\
 U Sgr & 6.745 & $1.605 \pm 0.023$ & 0.85 & $0.408 \pm 0.007$ & F95 & $0.08 \pm 0.08$ & G14b \\
 UU Mus & 11.636 & $0.306 \pm 0.012$ & 1.01 & $0.431 \pm 0.041$ & F95 & $0.11 \pm 0.09$ & G14b \\
 UX Car & 3.682 & $0.653 \pm 0.019$ & 1.02 & $0.102 \pm 0.023$ & F95 & $-0.10 \pm 0.15$ & G14b \\
 UX Per & 4.566 & $0.162 \pm 0.020$ & 1.17 & $0.462 \pm 0.024$ & F95 & $-0.05 \pm 0.06$ & G14b \\
 UY Car & 5.544 & $0.455 \pm 0.014$ & 0.94 & $0.188 \pm 0.017$ & F95 & $0.13 \pm 0.06$ & G14b \\
 UY Per & 5.365 & $0.415 \pm 0.015$ & 1.17 & $0.888 \pm 0.013$ & F95 & $0.18 \pm 0.06$ & G14b \\
 UZ Car & 5.205 & $0.401 \pm 0.013$ & 0.95 & $0.213 \pm 0.034$ & F95 & $0.13 \pm 0.06$ & G14b \\
 UZ Cas & 4.259 & $0.251 \pm 0.020$ & 1.23 & $0.469 \pm 0.034$ & F95 & $-0.05 \pm 0.06$ & G14b \\
 UZ Sct & 14.744 & $0.324 \pm 0.025$ & 0.91 & $0.959 \pm 0.023$ & F95 & $0.33 \pm 0.08$ & G15 \\
 V Car & 6.697 & $0.797 \pm 0.014$ & 1.04 & $0.164 \pm 0.013$ & F95 & $-0.06 \pm 0.15$ & G14b \\
 V Cen & 5.494 & $1.409 \pm 0.022$ & 1.06 & $0.265 \pm 0.016$ & F95 & $0.04 \pm 0.09$ & G14b \\
 V Lac & 4.983 & $0.496 \pm 0.016$ & 1.09 & $0.293 \pm 0.034$ & F95 & $0.06 \pm 0.06$ & G14b \\
 V Vel & 4.371 & $0.953 \pm 0.017$ & 1.03 & $0.225 \pm 0.021$ & F95 & $-0.30 \pm 0.15$ & G14b \\
 V0339 Cen & 9.466 & $0.568 \pm 0.021$ & 0.89 & $0.426 \pm 0.016$ & F95 & $0.06 \pm 0.03$ & G15 \\
 V0340 Ara & 20.814 & $0.239 \pm 0.020$ & 0.93 & $0.548 \pm 0.008$ & F95 & $0.33 \pm 0.09$ & G15 \\
 V0340 Nor & 11.289 & $0.491 \pm 0.025$ & 0.92 & $0.312 \pm 0.050$ & F95 & $0.07 \pm 0.07$ & G15 \\
 V0378 Cen & 6.460 & $0.524 \pm 0.019$ & 0.99 & $0.374 \pm 0.049$ & F95 & $0.08 \pm 0.06$ & G14b \\
 V0381 Cen & 5.079 & $0.818 \pm 0.020$ & 1.06 & $0.206 \pm 0.013$ & F95 & $0.02 \pm 0.06$ & G14b \\
 V0386 Cyg & 5.258 & $0.894 \pm 0.013$ & 0.95 & $0.907 \pm 0.033$ & F95 & $0.11 \pm 0.08$ & G14b \\
 V0402 Cyg & 4.365 & $0.410 \pm 0.011$ & 0.92 & $0.455 \pm 0.062$ & F95 & $0.02 \pm 0.08$ & G14b \\
 V0459 Cyg & 7.251 & $0.382 \pm 0.014$ & 1.09 & $0.775 \pm 0.024$ & F95 & $0.09 \pm 0.06$ & G14b \\
 V0470 Sco & 16.261 & $0.534 \pm 0.029$ & 0.97 & $1.550 \pm 0.124$ & F95 & $0.16 \pm 0.06$ & G15 \\
 V0493 Aql & 2.988 & $0.472 \pm 0.017$ & 1.12 & $0.730 \pm 0.087$ & F95 & $0.03 \pm 0.06$ & G14b \\
 V0496 Cen & 4.424 & $0.563 \pm 0.013$ & 0.94 & $0.579 \pm 0.031$ & F95 & $0.09 \pm 0.06$ & G14b \\
 V0520 Cyg & 4.049 & $0.437 \pm 0.012$ & 1.03 & $0.754 \pm 0.075$ & F95 & $0.08 \pm 0.06$ & G14b \\
 V0538 Cyg & 6.119 & $0.394 \pm 0.017$ & 0.99 & $0.656 \pm 0.021$ & F95 & $0.05 \pm 0.06$ & G14b \\
 V0600 Aql & 7.239 & $0.523 \pm 0.019$ & 1.13 & $0.812 \pm 0.007$ & F95 & $0.03 \pm 0.08$ & G14b \\
 V0609 Cyg & 31.088 & $0.295 \pm 0.017$ & 1.12 & $1.243 \pm 0.124$ & F95 & $0.22 \pm 0.06$ & G14b \\
 V0636 Cas & 8.377 & $1.372 \pm 0.018$ & 1.02 & $0.593 \pm 0.065$ & F95 & $0.07 \pm 0.08$ & G14b \\
 V0636 Sco & 6.797 & $1.180 \pm 0.034$ & 1.15 & $0.227 \pm 0.017$ & F95 & $0.10 \pm 0.06$ & G14b \\
 V0733 Aql & 6.179 & $0.244 \pm 0.015$ & 0.98 & $0.106 \pm 0.011$ & A12 & $0.08 \pm 0.08$ & G14b \\
 V0737 Cen & 7.066 & $1.213 \pm 0.019$ & 0.92 & $0.227 \pm 0.022$ & F95 & $0.14 \pm 0.06$ & G14b \\
 V1154 Cyg & 4.925 & $0.442 \pm 0.012$ & 1.04 & $0.315 \pm 0.031$ & F95 & $-0.10 \pm 0.08$ & G14b \\
 V1162 Aql & 5.376 & $0.823 \pm 0.023$ & 0.95 & $0.184 \pm 0.011$ & F95 & $0.01 \pm 0.08$ & G14b \\
 VW Cen & 15.036 & $0.260 \pm 0.016$ & 1.06 & $0.424 \pm 0.022$ & F95 & $0.41 \pm 0.08$ & G15 \\
 VW Cru & 5.265 & $0.738 \pm 0.016$ & 0.85 & $0.640 \pm 0.046$ & F95 & $0.19 \pm 0.06$ & G14b \\
 VY Car & 18.890 & $0.565 \pm 0.017$ & 0.92 & $0.270 \pm 0.019$ & F95 & $-0.06 \pm 0.15$ & G14b \\
 VY Cyg & 7.857 & $0.485 \pm 0.012$ & 1.07 & $0.596 \pm 0.021$ & F95 & $0.00 \pm 0.08$ & G14b \\
 VY Per & 5.532 & $0.485 \pm 0.017$ & 1.15 & $0.948 \pm 0.018$ & F95 & $0.04 \pm 0.06$ & G14b \\
 VY Sgr & 13.557 & $0.412 \pm 0.025$ & 0.81 & $0.903 \pm 0.243$ & F95 & $0.33 \pm 0.12$ & G15 \\
 VZ Cyg & 4.864 & $0.545 \pm 0.016$ & 1.31 & $0.291 \pm 0.015$ & F95 & $0.05 \pm 0.08$ & G14b \\
 VZ Pup & 23.175 & $0.220 \pm 0.015$ & 1.24 & $0.433 \pm 0.018$ & F95 & $-0.01 \pm 0.04$ & G15 \\
 W Gem & 7.914 & $1.006 \pm 0.028$ & 1.23 & $0.264 \pm 0.011$ & F95 & $0.02 \pm 0.06$ & G14b \\
 WW Pup & 5.517 & $0.212 \pm 0.016$ & 1.14 & $0.334 \pm 0.017$ & F95 & $0.13 \pm 0.16$ & G15 \\
 WX Pup & 8.937 & $0.387 \pm 0.015$ & 1.06 & $0.306 \pm 0.018$ & F95 & $-0.15 \pm 0.15$ & G14b \\
 WY Pup & 5.251 & $0.258 \pm 0.013$ & 1.02 & $0.259 \pm 0.031$ & F95 & $-0.10 \pm 0.08$ & G15 \\
 WZ Pup & 5.027 & $0.281 \pm 0.017$ & 1.35 & $0.196 \pm 0.022$ & F95 & $-0.07 \pm 0.06$ & G15 \\
 WZ Sgr & 21.851 & $0.612 \pm 0.028$ & 0.94 & $0.457 \pm 0.025$ & F95 & $0.28 \pm 0.08$ & G15 \\
 X Cru & 6.220 & $0.654 \pm 0.019$ & 0.95 & $0.294 \pm 0.019$ & F95 & $0.15 \pm 0.06$ & G14b \\
 X Cyg & 16.386 & $0.910 \pm 0.020$ & 1.28 & $0.251 \pm 0.010$ & F95 & $0.10 \pm 0.08$ & G14b \\
 X Pup & 25.973 & $0.397 \pm 0.020$ & 1.04 & $0.396 \pm 0.015$ & F95 & $0.02 \pm 0.08$ & G15 \\
 X Sct & 4.198 & $0.634 \pm 0.019$ & 0.80 & $0.581 \pm 0.030$ & F95 & $0.12 \pm 0.09$ & G15 \\
 X Sgr & 7.013 & $2.843 \pm 0.141$ & 1.22 & $0.189 \pm 0.020$ & F95 & $-0.21 \pm 0.30$ & G14b \\
 X Vul & 6.320 & $0.864 \pm 0.022$ & 1.06 & $0.775 \pm 0.021$ & F95 & $0.07 \pm 0.08$ & G14b \\
 XX Cen & 10.953 & $0.570 \pm 0.026$ & 1.24 & $0.245 \pm 0.012$ & F95 & $0.04 \pm 0.09$ & G14b \\
 XX Mon & 5.456 & $0.242 \pm 0.013$ & 0.86 & $0.586 \pm 0.014$ & F95 & $0.01 \pm 0.08$ & G15 \\
 XX Sgr & 6.424 & $0.724 \pm 0.027$ & 1.10 & $0.493 \pm 0.016$ & F95 & $-0.01 \pm 0.06$ & G15 \\
 XX Vel & 6.985 & $0.308 \pm 0.013$ & 0.88 & $0.530 \pm 0.007$ & F95 & $0.11 \pm 0.06$ & G14b \\
 XZ Car & 16.651 & $0.473 \pm 0.018$ & 1.05 & $0.372 \pm 0.026$ & F95 & $0.19 \pm 0.06$ & G14b \\
 Y Aur & 3.859 & $0.541 \pm 0.017$ & 1.12 & $0.384 \pm 0.031$ & F95 & $-0.26 \pm 0.15$ & G14b \\
 Y Lac & 4.324 & $0.431 \pm 0.013$ & 1.05 & $0.212 \pm 0.020$ & F95 & $0.03 \pm 0.06$ & G14b \\
 Y Oph & 17.125 & $1.348 \pm 0.036$ & 1.03 & $0.606 \pm 0.030$ & F95 & $0.06 \pm 0.08$ & G14b \\
 Y Sct & 10.342 & $0.558 \pm 0.020$ & 0.94 & $0.792 \pm 0.021$ & F95 & $0.23 \pm 0.06$ & G14b \\
 YZ Aur & 18.193 & $0.233 \pm 0.016$ & 0.99 & $0.548 \pm 0.055$ & F95 & $-0.33 \pm 0.15$ & G14b \\
 YZ Car & 18.168 & $0.358 \pm 0.018$ & 1.17 & $0.324 \pm 0.039$ & F95 & $0.00 \pm 0.06$ & G14b \\
 YZ Sgr & 9.554 & $0.860 \pm 0.024$ & 0.95 & $0.289 \pm 0.007$ & F95 & $0.06 \pm 0.08$ & G14b \\
 Z Lac & 10.886 & $0.510 \pm 0.021$ & 1.05 & $0.352 \pm 0.015$ & F95 & $0.10 \pm 0.06$ & G14b \\
 Z Sct & 12.901 & $0.357 \pm 0.018$ & 0.90 & $0.535 \pm 0.039$ & F95 & $0.12 \pm 0.09$ & G15 \\
  \hline
\label{table:data_MW_parameters}
\end{longtable*}

\newpage
 
\begin{longtable*}{l c c c c c c c}
\caption{Optical and NIR mean apparent magnitudes for the sample of Milky Way Cepheids. The magnitudes do not include the reddening correction. The uncertainties are only the random errors and do not include photometric zero point errors. \textbf{References}: (B97): \citet{Barnes1997}; (F08): \citet{Feast2008}; (L92): \citet{Laney1992}; (M11): \citet{Monson2011}; (W84): \citet{Welch1984}. All magnitudes in $V$ and $I$ are from \citet{Berdnikov2008}.}\\
\hline
\hline
Star	& $V$	& $I$		& $J$	& $H$	& $K_S$	& Ref. NIR \\
	&  (mag)	&  (mag)	&  (mag)	&  (mag)	& (mag)	&	   \\ 
\hline
\endfirsthead
\multicolumn{8}{c}{\textbf{Table \ref{table:data_MW_photometry}} (continued)} \\
\hline
Star	& $V$	& $I$		& $J$	& $H$	& $K_S$	& Ref. NIR \\
	&  (mag)	&  (mag)	&  (mag)	&  (mag)	& (mag)	&	   \\ 
\hline 
\endhead
\hline
\endfoot
\endlastfoot
 AA Gem & $9.735 \pm 0.008$ & --- & $7.647 \pm 0.011$ & $7.206 \pm 0.010$ & $7.069 \pm 0.020$& M11 \\
 AC Mon & $10.100 \pm 0.016$ & $8.708 \pm 0.012$ & $7.590 \pm 0.013$ & $7.072 \pm 0.011$ & $6.867 \pm 0.012$& M11 \\
 AD Gem & --- & --- & $8.453 \pm 0.006$ & $8.154 \pm 0.006$ & $8.043 \pm 0.007$& B97, M11 \\
 AD Pup & $9.898 \pm 0.006$ & $8.716 \pm 0.006$ & --- & --- & ---& --- \\
 AE Vel & $10.257 \pm 0.009$ & $8.730 \pm 0.007$ & --- & --- & ---& --- \\
 AG Cru & $8.228 \pm 0.006$ & $7.346 \pm 0.006$ & --- & --- & ---& --- \\
 AP Pup & $7.385 \pm 0.006$ & $6.463 \pm 0.006$ & --- & --- & ---& --- \\
 AP Sgr & $6.967 \pm 0.006$ & $6.053 \pm 0.006$ & --- & --- & ---& --- \\
 AQ Car & $8.892 \pm 0.023$ & $7.895 \pm 0.020$ & --- & --- & ---& --- \\
 AQ Pup & $8.690 \pm 0.006$ & $7.153 \pm 0.006$ & $6.000 \pm 0.006$ & $5.484 \pm 0.008$ & $5.256 \pm 0.009$& L92 \\
 AS Per & --- & --- & $6.941 \pm 0.014$ & $6.482 \pm 0.007$ & $6.279 \pm 0.017$& M11 \\
 AT Pup & $7.985 \pm 0.006$ & $7.080 \pm 0.006$ & --- & --- & ---& --- \\
 AV Sgr & $11.331 \pm 0.021$ & $8.851 \pm 0.013$ & --- & --- & ---& --- \\
 AW Per & $7.473 \pm 0.140$ & --- & $5.222 \pm 0.010$ & $4.836 \pm 0.009$ & $4.676 \pm 0.010$& M11 \\
 AX Cir & $5.887 \pm 0.006$ & $4.987 \pm 0.006$ & --- & --- & ---& --- \\
 AY Cas & $11.543 \pm 0.024$ & --- & --- & --- & ---& --- \\
 AY Cen & $8.818 \pm 0.006$ & $7.701 \pm 0.006$ & --- & --- & ---& --- \\
 AY Sgr & $10.559 \pm 0.012$ & $8.729 \pm 0.009$ & $7.140 \pm 0.008$ & $6.534 \pm 0.010$ & $6.282 \pm 0.016$& M11 \\
 BB Her & $10.093 \pm 0.007$ & $8.941 \pm 0.010$ & --- & --- & ---& --- \\
 BB Sgr & $6.952 \pm 0.006$ & $5.848 \pm 0.006$ & $5.025 \pm 0.006$ & $4.643 \pm 0.007$ & $4.496 \pm 0.008$& L92, W84 \\
 BE Mon & $10.574 \pm 0.008$ & $9.243 \pm 0.008$ & $8.265 \pm 0.014$ & $7.857 \pm 0.011$ & $7.701 \pm 0.024$& M11 \\
 BF Oph & $7.342 \pm 0.006$ & $6.367 \pm 0.006$ & $5.626 \pm 0.008$ & $5.298 \pm 0.007$ & $5.147 \pm 0.009$& L92, W84 \\
 BG Lac & $8.897 \pm 0.006$ & $7.811 \pm 0.014$ & $7.023 \pm 0.006$ & $6.655 \pm 0.006$ & $6.500 \pm 0.006$& B97, M11 \\
 BG Vel & $7.653 \pm 0.006$ & $6.342 \pm 0.006$ & --- & --- & ---& --- \\
 BK Aur & $9.445 \pm 0.015$ & --- & $7.300 \pm 0.015$ & $6.890 \pm 0.019$ & $6.735 \pm 0.021$& M11 \\
 BM Per & $10.428 \pm 0.028$ & --- & $6.680 \pm 0.015$ & $6.007 \pm 0.012$ & $5.724 \pm 0.018$& M11 \\
 BM Pup & $10.846 \pm 0.006$ & $9.414 \pm 0.006$ & --- & --- & ---& --- \\
 BN Pup & $9.907 \pm 0.016$ & $8.585 \pm 0.020$ & $7.534 \pm 0.008$ & $7.079 \pm 0.009$ & $6.880 \pm 0.008$& L92 \\
 BP Cas & $10.951 \pm 0.021$ & --- & --- & --- & ---& --- \\
 BZ Cyg & $10.221 \pm 0.006$ & $8.327 \pm 0.018$ & $6.774 \pm 0.014$ & $6.153 \pm 0.010$ & $5.879 \pm 0.014$& M11 \\
 $\beta$ Dor & $3.737 \pm 0.006$ & $2.939 \pm 0.006$ & $2.365 \pm 0.006$ & $2.038 \pm 0.006$ & $1.925 \pm 0.006$& F08, L92 \\
 CD Cas & $10.782 \pm 0.009$ & --- & $7.644 \pm 0.006$ & $7.093 \pm 0.012$ & $6.878 \pm 0.012$& M11 \\
 CD Cyg & $8.963 \pm 0.009$ & $7.498 \pm 0.028$ & $6.363 \pm 0.015$ & $5.853 \pm 0.012$ & $5.668 \pm 0.011$& W84, M11 \\
 CE Pup & $11.832 \pm 0.010$ & $9.968 \pm 0.007$ & --- & --- & ---& --- \\
 CF Cas & $11.138 \pm 0.006$ & $9.756 \pm 0.012$ & $8.606 \pm 0.010$ & $8.136 \pm 0.012$ & $7.923 \pm 0.012$& W84, M11 \\
 CG Cas & $11.378 \pm 0.010$ & --- & --- & --- & ---& --- \\
 CK Sct & --- & --- & $7.393 \pm 0.006$ & $6.822 \pm 0.010$ & $6.610 \pm 0.014$& M11 \\
 CN Car & $10.684 \pm 0.008$ & $9.355 \pm 0.009$ & --- & --- & ---& --- \\
 CP Cep & $10.588 \pm 0.012$ & $8.766 \pm 0.024$ & $7.348 \pm 0.010$ & $6.726 \pm 0.012$ & $6.492 \pm 0.012$& M11 \\
 CR Cep & $9.646 \pm 0.008$ & $7.979 \pm 0.020$ & $6.654 \pm 0.006$ & $6.101 \pm 0.007$ & $5.890 \pm 0.007$& M11 \\
 CR Ser & $10.857 \pm 0.009$ & $8.899 \pm 0.026$ & $7.353 \pm 0.007$ & $6.763 \pm 0.007$ & $6.503 \pm 0.012$& M11 \\
 CS Mon & $11.005 \pm 0.006$ & $9.651 \pm 0.006$ & --- & --- & ---& --- \\
 CS Ori & $11.399 \pm 0.037$ & $10.261 \pm 0.019$ & $9.341 \pm 0.011$ & $8.960 \pm 0.009$ & $8.810 \pm 0.017$& M11 \\
 CS Vel & $11.702 \pm 0.007$ & $10.069 \pm 0.007$ & $8.735 \pm 0.010$ & $8.228 \pm 0.014$ & $7.973 \pm 0.011$& L92, W84 \\
 CV Mon & $10.291 \pm 0.006$ & $8.645 \pm 0.006$ & $7.323 \pm 0.011$ & $6.790 \pm 0.007$ & $6.545 \pm 0.007$& L92, W84, M11 \\
 CY Car & $9.755 \pm 0.007$ & $8.712 \pm 0.006$ & --- & --- & ---& --- \\
 CY Cas & $11.643 \pm 0.020$ & --- & $7.876 \pm 0.028$ & $7.180 \pm 0.018$ & $6.915 \pm 0.023$& M11 \\
 CZ Cas & $11.752 \pm 0.009$ & $10.059 \pm 0.021$ & --- & --- & ---& --- \\
 DD Cas & $9.888 \pm 0.007$ & $8.561 \pm 0.025$ & $7.537 \pm 0.008$ & $7.073 \pm 0.011$ & $6.909 \pm 0.014$& M11 \\
 DF Cas & $10.879 \pm 0.006$ & --- & --- & --- & ---& --- \\
 DL Cas & $8.971 \pm 0.006$ & --- & $6.560 \pm 0.014$ & $6.106 \pm 0.011$ & $5.912 \pm 0.015$& W84, M11 \\
 DW Per & $11.577 \pm 0.008$ & --- & --- & --- & ---& --- \\
 $\delta$ Cep & $3.930 \pm 0.010$ & --- & $2.676 \pm 0.006$ & $2.393 \pm 0.006$ & $2.291 \pm 0.006$& F08, B97 \\
 EK Mon & $11.062 \pm 0.006$ & $9.617 \pm 0.006$ & --- & --- & ---& --- \\
 ER Car & $6.828 \pm 0.006$ & $5.961 \pm 0.006$ & --- & --- & ---& --- \\
 EX Vel & $11.573 \pm 0.007$ & $9.775 \pm 0.006$ & --- & --- & ---& --- \\
 EY Car & $10.359 \pm 0.010$ & $9.260 \pm 0.008$ & --- & --- & ---& --- \\
 Eta Aql & $3.878 \pm 0.006$ & $3.024 \pm 0.006$ & $2.386 \pm 0.006$ & $2.067 \pm 0.006$ & $1.951 \pm 0.006$& B97, W84 \\
 FI Car & $11.626 \pm 0.010$ & $9.855 \pm 0.009$ & --- & --- & ---& --- \\
 FM Aql & $8.278 \pm 0.006$ & $6.780 \pm 0.010$ & $5.681 \pm 0.006$ & $5.217 \pm 0.006$ & $5.026 \pm 0.006$& B97, W84, M11 \\
 FN Aql & $8.383 \pm 0.006$ & $6.992 \pm 0.006$ & $5.965 \pm 0.006$ & $5.495 \pm 0.006$ & $5.315 \pm 0.006$& B97, W84, M11 \\
 FN Vel & $10.303 \pm 0.006$ & $8.830 \pm 0.007$ & --- & --- & ---& --- \\
 GH Cyg & $9.904 \pm 0.006$ & $8.432 \pm 0.011$ & $7.262 \pm 0.011$ & $6.804 \pm 0.006$ & $6.598 \pm 0.016$& M11 \\
 GI Cyg & $11.745 \pm 0.012$ & --- & --- & --- & ---& --- \\
 GQ Ori & $8.965 \pm 0.011$ & $7.885 \pm 0.007$ & --- & --- & ---& --- \\
 GU Nor & $10.354 \pm 0.006$ & $8.799 \pm 0.007$ & --- & --- & ---& --- \\
 GX Car & $9.344 \pm 0.009$ & $8.137 \pm 0.006$ & --- & --- & ---& --- \\
 GY Sge & $10.163 \pm 0.006$ & --- & $5.604 \pm 0.008$ & $4.887 \pm 0.007$ & $4.546 \pm 0.006$& L92, W84 \\
 HW Car & $9.136 \pm 0.006$ & $8.028 \pm 0.006$ & --- & --- & ---& --- \\
 IQ Nor & $9.665 \pm 0.019$ & $8.115 \pm 0.020$ & --- & --- & ---& --- \\
 IT Car & $8.102 \pm 0.006$ & $7.070 \pm 0.006$ & --- & --- & ---& --- \\
 KK Cen & $11.452 \pm 0.036$ & $9.934 \pm 0.026$ & --- & --- & ---& --- \\
 KN Cen & $9.865 \pm 0.006$ & $7.994 \pm 0.006$ & $6.399 \pm 0.007$ & $5.747 \pm 0.008$ & $5.440 \pm 0.006$& L92 \\
 KQ Sco & $9.835 \pm 0.006$ & $7.659 \pm 0.006$ & $5.909 \pm 0.012$ & $5.215 \pm 0.010$ & $4.901 \pm 0.013$& L92, W84 \\
  $\ell$ Car & $3.723 \pm 0.006$ & $2.554 \pm 0.006$ & $1.679 \pm 0.006$ & $1.218 \pm 0.006$ & $1.054 \pm 0.006$& L92 \\
LS Pup & $10.462 \pm 0.007$ & $9.073 \pm 0.008$ & $7.999 \pm 0.006$ & $7.521 \pm 0.007$ & $7.312 \pm 0.006$& L92 \\
 MW Cyg & $9.483 \pm 0.006$ & --- & $6.700 \pm 0.006$ & $6.209 \pm 0.009$ & $5.998 \pm 0.014$& M11 \\
 MZ Cen & $11.553 \pm 0.007$ & $9.786 \pm 0.010$ & --- & --- & ---& --- \\
 QY Cen & $11.784 \pm 0.006$ & $9.350 \pm 0.007$ & --- & --- & ---& --- \\
 R Cru & $6.771 \pm 0.006$ & $5.901 \pm 0.006$ & --- & --- & ---& --- \\
 R Mus & $6.313 \pm 0.006$ & $5.497 \pm 0.006$ & --- & --- & ---& --- \\
 R TrA & $6.656 \pm 0.006$ & $5.843 \pm 0.006$ & --- & --- & ---& --- \\
 RR Lac & $8.846 \pm 0.006$ & $7.814 \pm 0.015$ & $6.977 \pm 0.008$ & $6.628 \pm 0.010$ & $6.488 \pm 0.011$& M11 \\
 RS Nor & $10.019 \pm 0.018$ & $8.541 \pm 0.013$ & --- & --- & ---& --- \\
 RS Ori & $8.410 \pm 0.011$ & $7.282 \pm 0.012$ & $6.408 \pm 0.016$ & $6.027 \pm 0.017$ & $5.880 \pm 0.019$& M11 \\
 RS Pup & $7.008 \pm 0.006$ & $5.478 \pm 0.006$ & $4.341 \pm 0.009$ & $3.830 \pm 0.007$ & $3.605 \pm 0.008$& L92, W84 \\
 RT Aur & $5.469 \pm 0.076$ & $4.811 \pm 0.043$ & $4.236 \pm 0.008$ & $3.998 \pm 0.007$ & $3.906 \pm 0.006$& B97, M11 \\
 RU Sct & --- & --- & $5.909 \pm 0.008$ & $5.298 \pm 0.007$ & $5.036 \pm 0.009$& L92, W84, M11 \\
 RV Sco & $7.046 \pm 0.006$ & $5.907 \pm 0.006$ & --- & --- & ---& --- \\
 RW Cam & $8.657 \pm 0.010$ & --- & $5.828 \pm 0.012$ & $5.291 \pm 0.013$ & $5.093 \pm 0.010$& M11 \\
 RW Cas & $9.248 \pm 0.019$ & $7.871 \pm 0.020$ & $6.841 \pm 0.024$ & $6.372 \pm 0.011$ & $6.194 \pm 0.026$& M11 \\
 RX Aur & $7.670 \pm 0.007$ & --- & $5.737 \pm 0.008$ & $5.363 \pm 0.011$ & $5.233 \pm 0.017$& M11 \\
 RX Cam & $7.670 \pm 0.012$ & --- & $5.178 \pm 0.023$ & $4.732 \pm 0.020$ & $4.561 \pm 0.012$& M11 \\
 RY CMa & $8.109 \pm 0.006$ & $7.133 \pm 0.006$ & --- & --- & ---& --- \\
 RY Sco & $7.999 \pm 0.006$ & $6.253 \pm 0.006$ & $4.899 \pm 0.006$ & $4.368 \pm 0.006$ & $4.102 \pm 0.007$& L92, W84 \\
 RY Vel & $8.376 \pm 0.006$ & $6.827 \pm 0.006$ & $5.604 \pm 0.008$ & $5.124 \pm 0.007$ & $4.886 \pm 0.006$& L92, W84 \\
 RZ CMa & $9.702 \pm 0.007$ & $8.504 \pm 0.007$ & --- & --- & ---& --- \\
 RZ Gem & $10.048 \pm 0.249$ & --- & $7.612 \pm 0.010$ & $7.169 \pm 0.009$ & $6.970 \pm 0.015$& M11 \\
 RZ Vel & $7.089 \pm 0.006$ & $5.862 \pm 0.006$ & $4.889 \pm 0.012$ & $4.463 \pm 0.007$ & $4.267 \pm 0.006$& L92 \\
 S Cru & $6.601 \pm 0.006$ & $5.732 \pm 0.006$ & --- & --- & ---& --- \\
 S Mus & $6.133 \pm 0.006$ & $5.199 \pm 0.006$ & $4.473 \pm 0.006$ & $4.135 \pm 0.006$ & $3.983 \pm 0.008$& L92, W84 \\
 S Nor & $6.427 \pm 0.006$ & $5.428 \pm 0.006$ & $4.652 \pm 0.006$ & $4.286 \pm 0.006$ & $4.131 \pm 0.008$& L92, W84 \\
 S Sge & $5.612 \pm 0.006$ & $4.772 \pm 0.010$ & $4.155 \pm 0.006$ & $3.847 \pm 0.006$ & $3.732 \pm 0.006$& W84, B97 \\
 S TrA & $6.391 \pm 0.006$ & $5.592 \pm 0.006$ & --- & --- & ---& --- \\
 SS CMa & --- & $8.480 \pm 0.010$ & --- & --- & ---& --- \\
 SS Sct & --- & --- & $6.299 \pm 0.008$ & $5.938 \pm 0.006$ & $5.807 \pm 0.008$& W84, M11 \\
 ST Tau & $8.243 \pm 0.014$ & $7.171 \pm 0.016$ & --- & --- & ---& --- \\
 ST Vel & $9.699 \pm 0.006$ & $8.286 \pm 0.006$ & --- & --- & ---& --- \\
 SU Cyg & $6.855 \pm 0.007$ & $6.198 \pm 0.013$ & $5.638 \pm 0.007$ & $5.397 \pm 0.007$ & $5.295 \pm 0.008$& W84, M11 \\
 SV Mon & $8.266 \pm 0.008$ & $7.139 \pm 0.006$ & $6.262 \pm 0.015$ & $5.835 \pm 0.010$ & $5.691 \pm 0.017$& M11 \\
 SV Per & $8.977 \pm 0.011$ & --- & $6.802 \pm 0.021$ & $6.360 \pm 0.016$ & $6.198 \pm 0.018$& M11 \\
 SV Vel & $8.583 \pm 0.006$ & $7.329 \pm 0.006$ & --- & --- & ---& --- \\
 SV Vul & $7.216 \pm 0.006$ & $5.697 \pm 0.009$ & $4.571 \pm 0.006$ & $4.077 \pm 0.006$ & $3.887 \pm 0.006$& W84, L92, B97, M11 \\
 SW Cas & $9.713 \pm 0.007$ & $8.438 \pm 0.020$ & $7.412 \pm 0.009$ & $6.987 \pm 0.013$ & $6.820 \pm 0.015$& M11 \\
 SW Vel & $8.137 \pm 0.014$ & $6.850 \pm 0.008$ & $5.852 \pm 0.018$ & $5.407 \pm 0.012$ & $5.203 \pm 0.011$& L92 \\
 SX Car & $9.082 \pm 0.006$ & $8.039 \pm 0.006$ & --- & --- & ---& --- \\
 SX Per & $11.223 \pm 0.104$ & --- & $8.769 \pm 0.010$ & $8.352 \pm 0.007$ & $8.187 \pm 0.013$& M11 \\
 SX Vel & $8.278 \pm 0.006$ & $7.262 \pm 0.006$ & $6.474 \pm 0.006$ & $6.127 \pm 0.006$ & $5.965 \pm 0.006$& L92 \\
 SY Aur & $9.069 \pm 0.009$ & --- & $6.923 \pm 0.009$ & $6.530 \pm 0.012$ & $6.367 \pm 0.014$& M11 \\
 SY Nor & $9.520 \pm 0.023$ & $7.949 \pm 0.030$ & --- & --- & ---& --- \\
 SZ Aql & $8.636 \pm 0.011$ & $7.082 \pm 0.015$ & $5.865 \pm 0.008$ & $5.351 \pm 0.006$ & $5.138 \pm 0.006$& B97, W84, L92, M11 \\
 SZ Cas & $9.843 \pm 0.006$ & $8.110 \pm 0.008$ & --- & --- & ---& --- \\
 SZ Cyg & $9.435 \pm 0.011$ & $7.798 \pm 0.026$ & $6.530 \pm 0.009$ & $5.960 \pm 0.007$ & $5.732 \pm 0.014$& M11 \\
 T Ant & $9.331 \pm 0.006$ & $8.523 \pm 0.006$ & --- & --- & ---& --- \\
 T Cru & $6.570 \pm 0.006$ & $5.608 \pm 0.006$ & --- & --- & ---& --- \\
 T Mon & $6.138 \pm 0.006$ & $4.987 \pm 0.006$ & $4.092 \pm 0.009$ & $3.648 \pm 0.009$ & $3.487 \pm 0.008$& W84, L92, M11 \\
 T Vel & $8.029 \pm 0.006$ & $6.964 \pm 0.006$ & $6.143 \pm 0.006$ & $5.775 \pm 0.006$ & $5.605 \pm 0.006$& L92 \\
 T Vul & $5.750 \pm 0.006$ & $5.077 \pm 0.015$ & $4.532 \pm 0.006$ & $4.272 \pm 0.006$ & $4.174 \pm 0.006$& W84, B97 \\
 TT Aql & $7.141 \pm 0.006$ & $5.732 \pm 0.009$ & $4.671 \pm 0.009$ & $4.194 \pm 0.007$ & $4.017 \pm 0.006$& W84, M11, B97 \\
 TV CMa & $10.587 \pm 0.011$ & $9.173 \pm 0.011$ & $8.035 \pm 0.008$ & $7.588 \pm 0.011$ & $7.386 \pm 0.014$& M11 \\
 TV Cam & $11.729 \pm 0.018$ & --- & --- & --- & ---& --- \\
 TW CMa & $9.573 \pm 0.007$ & $8.458 \pm 0.007$ & $7.577 \pm 0.010$ & $7.183 \pm 0.009$ & $7.029 \pm 0.017$& M11 \\
 TW Nor & $11.670 \pm 0.007$ & $9.306 \pm 0.010$ & $7.406 \pm 0.022$ & $6.705 \pm 0.013$ & $6.358 \pm 0.029$& L92, W84 \\
 TX Cen & $10.527 \pm 0.006$ & $8.618 \pm 0.006$ & --- & --- & ---& --- \\
 TX Cyg & $9.490 \pm 0.012$ & $7.225 \pm 0.030$ & $5.342 \pm 0.020$ & $4.633 \pm 0.018$ & $4.323 \pm 0.018$& M11 \\
 TX Mon & $10.960 \pm 0.010$ & $9.634 \pm 0.008$ & $8.581 \pm 0.013$ & $8.121 \pm 0.013$ & $7.943 \pm 0.017$& M11 \\
 TY Sct & $10.821 \pm 0.013$ & $8.811 \pm 0.016$ & $7.247 \pm 0.011$ & $6.637 \pm 0.009$ & $6.386 \pm 0.021$& M11 \\
 TZ Mon & $10.793 \pm 0.008$ & $9.472 \pm 0.006$ & $8.458 \pm 0.012$ & $8.009 \pm 0.014$ & $7.815 \pm 0.016$& M11 \\
 TZ Mus & $11.690 \pm 0.006$ & $10.144 \pm 0.008$ & --- & --- & ---& --- \\
 U Aql & $6.432 \pm 0.006$ & $5.271 \pm 0.010$ & $4.389 \pm 0.012$ & $3.999 \pm 0.009$ & $3.844 \pm 0.010$& W84, M11 \\
 U Car & $6.284 \pm 0.006$ & $5.052 \pm 0.006$ & $4.104 \pm 0.007$ & $3.674 \pm 0.006$ & $3.483 \pm 0.006$& L92, W84 \\
 U Nor & --- & --- & $5.825 \pm 0.006$ & $5.236 \pm 0.006$ & $4.944 \pm 0.006$& L92 \\
 U Sgr & $6.697 \pm 0.006$ & $5.436 \pm 0.006$ & $4.512 \pm 0.006$ & $4.100 \pm 0.006$ & $3.933 \pm 0.007$& W84, L92, M11 \\
 U Vul & $7.122 \pm 0.006$ & $5.602 \pm 0.011$ & $4.528 \pm 0.009$ & $4.093 \pm 0.007$ & $3.912 \pm 0.006$& B97, M11 \\
 UU Mus & --- & --- & $7.439 \pm 0.007$ & $6.994 \pm 0.006$ & $6.788 \pm 0.006$& L92 \\
 UW Car & $9.424 \pm 0.008$ & $8.218 \pm 0.013$ & --- & --- & ---& --- \\
 UX Car & $8.295 \pm 0.006$ & $7.554 \pm 0.006$ & --- & --- & ---& --- \\
 UX Per & $11.650 \pm 0.018$ & --- & --- & --- & ---& --- \\
 UY Car & $8.948 \pm 0.006$ & $8.007 \pm 0.010$ & --- & --- & ---& --- \\
 UY Per & $11.319 \pm 0.013$ & $9.493 \pm 0.016$ & --- & --- & ---& --- \\
 UZ Car & $9.327 \pm 0.006$ & $8.365 \pm 0.006$ & --- & --- & ---& --- \\
 UZ Cas & $11.379 \pm 0.007$ & --- & --- & --- & ---& --- \\
 UZ Sct & $11.289 \pm 0.022$ & $9.148 \pm 0.035$ & $7.418 \pm 0.010$ & $6.741 \pm 0.010$ & $6.485 \pm 0.016$& M11 \\
 V Car & $7.368 \pm 0.006$ & $6.433 \pm 0.006$ & $5.728 \pm 0.006$ & $5.396 \pm 0.006$ & $5.249 \pm 0.006$& L92 \\
 V Cen & $6.830 \pm 0.006$ & $5.794 \pm 0.006$ & $4.995 \pm 0.006$ & $4.638 \pm 0.006$ & $4.479 \pm 0.009$& L92, W84 \\
 V Lac & $8.932 \pm 0.007$ & --- & --- & --- & ---& --- \\
 V Vel & $7.586 \pm 0.006$ & $6.691 \pm 0.006$ & --- & --- & ---& --- \\
 V0339 Cen & $8.714 \pm 0.013$ & $7.384 \pm 0.010$ & --- & --- & ---& --- \\
 V0340 Ara & $10.228 \pm 0.014$ & $8.580 \pm 0.007$ & --- & --- & ---& --- \\
 V0340 Nor & $8.403 \pm 0.008$ & $7.167 \pm 0.008$ & --- & --- & ---& --- \\
 V0350 Sgr & $7.481 \pm 0.006$ & $6.435 \pm 0.006$ & $5.627 \pm 0.011$ & $5.256 \pm 0.011$ & $5.130 \pm 0.008$& W84 \\
 V0378 Cen & $8.479 \pm 0.006$ & $7.260 \pm 0.006$ & --- & --- & ---& --- \\
 V0381 Cen & $7.675 \pm 0.006$ & $6.791 \pm 0.006$ & --- & --- & ---& --- \\
 V0386 Cyg & $9.624 \pm 0.007$ & --- & $6.375 \pm 0.007$ & $5.809 \pm 0.007$ & $5.540 \pm 0.015$& M11 \\
 V0395 Cas & $10.748 \pm 0.019$ & $9.447 \pm 0.035$ & --- & --- & ---& --- \\
 V0402 Cyg & $9.864 \pm 0.006$ & --- & $7.809 \pm 0.006$ & $7.416 \pm 0.006$ & $7.263 \pm 0.015$& M11 \\
 V0459 Cyg & $10.576 \pm 0.033$ & $8.881 \pm 0.019$ & $7.613 \pm 0.011$ & $7.075 \pm 0.010$ & $6.859 \pm 0.016$& M11 \\
 V0470 Sco & $11.005 \pm 0.008$ & $8.246 \pm 0.007$ & --- & --- & ---& --- \\
 V0493 Aql & $11.046 \pm 0.006$ & --- & --- & --- & ---& --- \\
 V0496 Aql & $7.769 \pm 0.006$ & $6.489 \pm 0.008$ & --- & --- & ---& --- \\
 V0496 Cen & $9.945 \pm 0.006$ & $8.539 \pm 0.006$ & --- & --- & ---& --- \\
 V0508 Mon & $10.502 \pm 0.006$ & $9.461 \pm 0.006$ & --- & --- & ---& --- \\
 V0520 Cyg & $10.852 \pm 0.006$ & $9.306 \pm 0.029$ & --- & --- & ---& --- \\
 V0538 Cyg & $10.449 \pm 0.009$ & $8.971 \pm 0.049$ & $7.803 \pm 0.007$ & $7.311 \pm 0.006$ & $7.119 \pm 0.008$& M11 \\
 V0600 Aql & $10.034 \pm 0.006$ & $8.281 \pm 0.011$ & --- & --- & ---& --- \\
 V0609 Cyg & $11.026 \pm 0.017$ & $8.683 \pm 0.015$ & $6.832 \pm 0.013$ & $6.128 \pm 0.010$ & $5.800 \pm 0.017$& M11 \\
 V0636 Cas & $7.183 \pm 0.006$ & --- & --- & --- & ---& --- \\
 V0636 Sco & $6.654 \pm 0.006$ & $5.649 \pm 0.006$ & --- & --- & ---& --- \\
 V0733 Aql & $9.976 \pm 0.006$ & $9.040 \pm 0.012$ & --- & --- & ---& --- \\
 V0737 Cen & $6.724 \pm 0.006$ & $5.701 \pm 0.006$ & --- & --- & ---& --- \\
 V1154 Cyg & $9.186 \pm 0.006$ & $8.180 \pm 0.018$ & --- & --- & ---& --- \\
 V1162 Aql & $7.806 \pm 0.006$ & $6.850 \pm 0.007$ & $6.143 \pm 0.008$ & $5.814 \pm 0.017$ & $5.682 \pm 0.020$& M11 \\
 VV Cas & $10.768 \pm 0.016$ & $9.432 \pm 0.018$ & $8.328 \pm 0.008$ & $7.885 \pm 0.006$ & $7.719 \pm 0.008$& M11 \\
 VW Cen & $10.263 \pm 0.007$ & $8.783 \pm 0.006$ & $7.555 \pm 0.007$ & $7.015 \pm 0.006$ & $6.775 \pm 0.006$& L92 \\
 VW Cru & $9.597 \pm 0.009$ & $7.977 \pm 0.006$ & --- & --- & ---& --- \\
 VW Pup & $11.393 \pm 0.007$ & $10.091 \pm 0.006$ & --- & --- & ---& --- \\
 VY Car & $7.460 \pm 0.006$ & $6.283 \pm 0.006$ & $5.375 \pm 0.015$ & $4.943 \pm 0.010$ & $4.760 \pm 0.010$& L92, W84 \\
 VY Cyg & $9.594 \pm 0.006$ & $8.127 \pm 0.019$ & $7.009 \pm 0.006$ & $6.552 \pm 0.009$ & $6.355 \pm 0.010$& M11 \\
 VY Per & $11.221 \pm 0.014$ & $9.297 \pm 0.026$ & --- & --- & ---& --- \\
 VY Sgr & $11.469 \pm 0.012$ & $9.129 \pm 0.013$ & --- & --- & ---& --- \\
 VZ Cyg & $8.970 \pm 0.008$ & $7.966 \pm 0.015$ & $7.201 \pm 0.007$ & $6.864 \pm 0.006$ & $6.721 \pm 0.006$& B97, W84, M11 \\
 VZ Pup & $9.657 \pm 0.009$ & $8.302 \pm 0.006$ & $7.277 \pm 0.007$ & $6.830 \pm 0.006$ & $6.626 \pm 0.006$& L92 \\
 W Gem & $7.012 \pm 0.049$ & --- & $5.129 \pm 0.033$ & $4.771 \pm 0.026$ & $4.656 \pm 0.021$& M11 \\
 W Sgr & $4.664 \pm 0.006$ & $3.850 \pm 0.006$ & --- & --- & ---& --- \\
 WW Car & $9.750 \pm 0.010$ & $8.644 \pm 0.007$ & --- & --- & ---& --- \\
 WW Pup & $10.599 \pm 0.010$ & $9.525 \pm 0.007$ & --- & --- & ---& --- \\
 WX Pup & $9.070 \pm 0.007$ & $7.968 \pm 0.006$ & --- & --- & ---& --- \\
 WY Pup & $10.599 \pm 0.013$ & $9.662 \pm 0.008$ & --- & --- & ---& --- \\
 WZ Pup & $10.320 \pm 0.006$ & $9.424 \pm 0.006$ & --- & --- & ---& --- \\
 WZ Sgr & $8.046 \pm 0.011$ & $6.544 \pm 0.010$ & $5.282 \pm 0.008$ & $4.761 \pm 0.006$ & $4.538 \pm 0.008$& L92, W84, M11 \\
 X Cru & $8.404 \pm 0.006$ & --- & --- & --- & ---& --- \\
 X Cyg & $6.385 \pm 0.009$ & $5.236 \pm 0.028$ & $4.383 \pm 0.008$ & $3.960 \pm 0.006$ & $3.799 \pm 0.006$& W84, B97 \\
 X Pup & $8.517 \pm 0.011$ & $7.161 \pm 0.006$ & $6.077 \pm 0.023$ & $5.599 \pm 0.011$ & $5.386 \pm 0.011$& L92 \\
 X Sct & $10.031 \pm 0.017$ & $8.613 \pm 0.033$ & --- & --- & ---& --- \\
 X Sgr & $4.548 \pm 0.006$ & $3.652 \pm 0.006$ & $2.950 \pm 0.007$ & $2.635 \pm 0.007$ & $2.505 \pm 0.010$& F08, W84 \\
 X Vul & $8.834 \pm 0.006$ & $7.198 \pm 0.020$ & $5.928 \pm 0.010$ & $5.433 \pm 0.008$ & $5.214 \pm 0.015$& M11 \\
 XX Cen & $7.824 \pm 0.006$ & $6.743 \pm 0.006$ & $5.914 \pm 0.008$ & $5.541 \pm 0.006$ & $5.375 \pm 0.007$& L92, W84 \\
 XX Mon & $11.914 \pm 0.007$ & $10.505 \pm 0.009$ & --- & --- & ---& --- \\
 XX Sgr & $8.869 \pm 0.006$ & $7.506 \pm 0.006$ & $6.412 \pm 0.033$ & $5.964 \pm 0.018$ & $5.799 \pm 0.022$& W84 \\
 XX Vel & $10.676 \pm 0.006$ & $9.302 \pm 0.006$ & --- & --- & ---& --- \\
 XZ Car & $8.597 \pm 0.006$ & $7.248 \pm 0.006$ & --- & --- & ---& --- \\
 Y Aur & $9.809 \pm 0.044$ & --- & $7.660 \pm 0.007$ & $7.291 \pm 0.008$ & $7.133 \pm 0.026$& M11 \\
 Y Lac & $9.159 \pm 0.007$ & $8.308 \pm 0.026$ & $7.626 \pm 0.006$ & $7.316 \pm 0.006$ & $7.201 \pm 0.008$& B97, M11 \\
 Y Oph & $6.148 \pm 0.006$ & $4.533 \pm 0.006$ & $3.349 \pm 0.006$ & $2.874 \pm 0.006$ & $2.662 \pm 0.008$& W84, L92 \\
 Y Sct & --- & --- & $6.472 \pm 0.009$ & $5.897 \pm 0.011$ & $5.646 \pm 0.014$& M11 \\
 Y Sgr & $5.739 \pm 0.006$ & $4.790 \pm 0.006$ & --- & --- & ---& --- \\
 YZ Aur & $10.346 \pm 0.009$ & --- & $7.498 \pm 0.015$ & $6.905 \pm 0.011$ & $6.689 \pm 0.024$& M11 \\
 YZ Car & $8.714 \pm 0.006$ & $7.438 \pm 0.006$ & --- & --- & ---& --- \\
 YZ Sgr & $7.351 \pm 0.006$ & $6.226 \pm 0.006$ & $5.379 \pm 0.007$ & $5.004 \pm 0.009$ & $4.861 \pm 0.010$& M11, W84 \\
 Z Lac & $8.417 \pm 0.006$ & $7.198 \pm 0.043$ & $6.235 \pm 0.009$ & $5.811 \pm 0.006$ & $5.653 \pm 0.008$& B97, M11 \\
 Z Sct & --- & --- & $6.962 \pm 0.017$ & $6.483 \pm 0.016$ & $6.282 \pm 0.017$& M11 \\
 $\zeta$ Gem & $3.889 \pm 0.006$ & $3.096 \pm 0.006$ & $2.538 \pm 0.006$ & $2.210 \pm 0.006$ & $2.096 \pm 0.006$& F08 \\
 \hline
\label{table:data_MW_photometry}
\end{longtable*}

\newpage

\begin{table*}[]
\footnotesize
\begin{center}
\caption{\new{Sample of Large Magellanic Cloud Cepheids and their main parameters. The Cepheid names in the first column are of the form OGLE-LMC-CEP-XXXX. The uncertainties on $V$ and $I$ band mean magnitudes are 0.02 mag and the uncertainty on $E(B-V)$ values is 0.017 mag. The distances listed in column (5) are corrected for their position in the LMC by the equations provided in Sect. \ref{sect:dist_LMC}. Apparent magnitudes in this table are not corrected for the reddening. The full table is available as supplementary material.}}
\begin{tabular}{c c c c c c c c c c c c c c c}
\hline
\hline
Cepheid & P & $\alpha$ & $\delta$ & $d$ & $V$ & $I$ & $J$ & $H$ & $K_S$ & $E(B-V)$  \\
 & (days) & (J2000) & (J2000) & (kpc) & (mag) & (mag) & (mag) & (mag) & (mag) & (mag)  \\
\hline
0107 & 8.739 & 72.214 & -69.356 & $50.63 \pm 0.56$ & 14.761 & 13.947 & $13.332 \pm 0.018$ & $13.027 \pm 0.015$ & $12.941 \pm 0.014$ & 0.182 \\ 
0174 & 15.863 & 72.719 & -69.316 & $50.55 \pm 0.56$ & 14.739 & 13.666 & $12.876 \pm 0.022$ & $12.454 \pm 0.020$ & $12.312 \pm 0.019$ & 0.174 \\ 
0328 & 34.460 & 73.599 & -70.902 & $50.68 \pm 0.56$ & 13.124 & 12.088 & $11.460 \pm 0.031$ & $11.111 \pm 0.027$ & $11.001 \pm 0.024$ & 0.133 \\ 
0467 & 22.718 & 74.301 & -67.383 & $50.06 \pm 0.55$ & 13.704 & 12.775 & $12.113 \pm 0.033$ & $11.767 \pm 0.031$ & $11.668 \pm 0.029$ & 0.112 \\ 
0473 & 2.634 & 74.331 & -68.821 & $50.25 \pm 0.56$ & 16.335 & 15.590 & $15.093 \pm 0.082$ & $14.856 \pm 0.116$ & $14.641 \pm 0.114$ & 0.141 \\ 
0478 & 2.764 & 74.355 & -69.567 & $50.36 \pm 0.56$ & 16.160 & 15.471 & $14.961 \pm 0.057$ & $14.666 \pm 0.073$ & $14.624 \pm 0.076$ & 0.150 \\ 
0480 & 4.035 & 74.364 & -69.355 & $50.33 \pm 0.56$ & 16.865 & 15.779 & $15.107 \pm 0.049$ & $14.513 \pm 0.052$ & $14.515 \pm 0.080$ & 0.161 \\ 
0482 & 7.466 & 74.370 & -69.227 & $50.31 \pm 0.56$ & 15.655 & 14.661 & $13.977 \pm 0.081$ & $13.439 \pm 0.031$ & $13.315 \pm 0.030$ & 0.151 \\ 
0487 & 3.109 & 74.422 & -69.406 & $50.33 \pm 0.56$ & 16.221 & 15.469 & $15.002 \pm 0.124$ & $14.614 \pm 0.041$ & $14.488 \pm 0.090$ & 0.165 \\ 
0488 & 3.647 & 74.422 & -68.800 & $50.24 \pm 0.56$ & 16.535 & 15.608 & $14.921 \pm 0.037$ & $14.517 \pm 0.045$ & $14.451 \pm 0.066$ & 0.140 \\ 
0494 & 2.727 & 74.441 & -69.062 & $50.27 \pm 0.56$ & 16.973 & 16.013 & $15.441 \pm 0.087$ & $14.953 \pm 0.100$ & $14.658 \pm 0.091$ & 0.134 \\ 
0498 & 3.630 & 74.455 & -68.720 & $50.22 \pm 0.56$ & 15.914 & 15.154 & $14.617 \pm 0.071$ & $14.295 \pm 0.059$ & $14.286 \pm 0.061$ & 0.139 \\ 
0514 & 3.504 & 74.554 & -69.203 & $50.28 \pm 0.56$ & 16.276 & 15.458 & $14.879 \pm 0.063$ & $14.458 \pm 0.055$ & $14.381 \pm 0.068$ & 0.152 \\ 
0518 & 3.249 & 74.577 & -69.367 & $50.30 \pm 0.56$ & 16.550 & 15.589 & $14.988 \pm 0.044$ & $14.565 \pm 0.044$ & $14.543 \pm 0.046$ & 0.174 \\ 
0529 & 2.856 & 74.637 & -69.041 & $50.24 \pm 0.56$ & 16.603 & 15.777 & $15.168 \pm 0.087$ & $14.746 \pm 0.084$ & $14.698 \pm 0.087$ & 0.134 \\ 
0539 & 3.455 & 74.672 & -68.865 & $50.21 \pm 0.55$ & 16.064 & 15.320 & $14.758 \pm 0.056$ & $14.373 \pm 0.069$ & $14.299 \pm 0.054$ & 0.131 \\ 
0540 & 3.750 & 74.673 & -69.526 & $50.31 \pm 0.56$ & 15.841 & 15.069 & $14.574 \pm 0.061$ & $14.198 \pm 0.058$ & $14.188 \pm 0.056$ & 0.160 \\ 
... & ... & ... & ... & ... & ... & ... & ... & ... & ... & ... \\
\hline
\end{tabular}
\flushleft
\label{table:data_LMC}
\end{center}
\end{table*}

\begin{table*}[]
\footnotesize
\begin{center}
\caption{\new{Sample of Small Magellanic Cloud Cepheids and their main parameters. The Cepheid names in the first column are of the form OGLE-SMC-CEP-XXXX. The uncertainties on $V$ and $I$ band mean magnitudes are 0.02 mag and the uncertainty on $E(B-V)$ values is 0.015 mag. The distances listed in column (5) are corrected for their position in the SMC by the equations provided in Sect. \ref{sect:dist_SMC}. Apparent magnitudes in this table are not corrected for the reddening. The full table is available as supplementary material. }}
\begin{tabular}{c c c c c c c c c c c c c c c}
\hline
\hline
Cepheid & P & $\alpha$ & $\delta$ & $d$ & $V$ & $I$ & $J$ & $K_S$ & $E(B-V)$  \\
 & (days) & (J2000) & (J2000) & (kpc) & (mag) & (mag) & (mag) & (mag) & (mag)  \\
\hline
0443 & 4.037 & 10.525 & -73.061 & $63.36 \pm 0.95$ & 16.443 & 15.671 & $15.139 \pm 0.018$ & $14.742 \pm 0.006$ & 0.096 \\ 
0489 & 3.242 & 10.734 & -73.160 & $63.42 \pm 0.95$ & 16.493 & 15.822 & $15.096 \pm 0.006$ & $14.492 \pm 0.010$ & 0.101 \\ 
0494 & 4.758 & 10.742 & -73.335 & $63.72 \pm 0.96$ & 15.884 & 15.025 & $14.447 \pm 0.006$ & $14.030 \pm 0.004$ & 0.103 \\ 
0495 & 6.312 & 10.743 & -73.092 & $63.30 \pm 0.95$ & 16.038 & 15.217 & $14.604 \pm 0.012$ & $14.164 \pm 0.004$ & 0.097 \\ 
0499 & 6.229 & 10.751 & -73.304 & $63.66 \pm 0.96$ & 15.986 & 15.258 & $14.740 \pm 0.004$ & $14.374 \pm 0.006$ & 0.108 \\ 
0514 & 2.542 & 10.774 & -73.082 & $63.27 \pm 0.95$ & 16.842 & 16.145 & $15.687 \pm 0.010$ & $15.306 \pm 0.010$ & 0.097 \\ 
0518 & 15.773 & 10.802 & -73.326 & $63.67 \pm 0.96$ & 15.184 & 14.173 & $13.449 \pm 0.002$ & $12.948 \pm 0.004$ & 0.108 \\ 
0524 & 10.527 & 10.828 & -73.339 & $63.68 \pm 0.96$ & 15.383 & 14.536 & $13.934 \pm 0.008$ & $13.492 \pm 0.006$ & 0.105 \\ 
0533 & 3.909 & 10.859 & -73.254 & $63.52 \pm 0.95$ & 16.122 & 15.390 & $14.895 \pm 0.006$ & $14.489 \pm 0.010$ & 0.105 \\ 
0551 & 3.262 & 10.905 & -73.129 & $63.28 \pm 0.95$ & 16.473 & 15.807 & $15.302 \pm 0.006$ & $14.954 \pm 0.016$ & 0.101 \\ 
0570 & 10.883 & 10.947 & -73.241 & $63.45 \pm 0.95$ & 15.213 & 14.354 & $13.738 \pm 0.010$ & $13.294 \pm 0.004$ & 0.112 \\ 
0571 & 4.897 & 10.948 & -73.335 & $63.62 \pm 0.95$ & 15.872 & 15.178 & $14.662 \pm 0.006$ & $14.297 \pm 0.008$ & 0.104 \\ 
0576 & 14.426 & 10.963 & -73.333 & $63.61 \pm 0.95$ & 15.110 & 14.122 & $13.470 \pm 0.016$ & $12.923 \pm 0.018$ & 0.104 \\ 
0584 & 4.654 & 10.989 & -73.192 & $63.35 \pm 0.95$ & 16.058 & 15.296 & $14.777 \pm 0.006$ & $14.418 \pm 0.006$ & 0.111 \\ 
0596 & 3.072 & 11.013 & -73.277 & $63.48 \pm 0.95$ & 17.170 & 16.282 & $15.673 \pm 0.006$ & $15.184 \pm 0.012$ & 0.120 \\ 
... & ... & ... & ... & ... & ... & ... & ... & ... & ... \\
\hline
\end{tabular}
\flushleft
\label{table:data_SMC}
\end{center}
\end{table*}

\end{document}